\documentclass[journal]{IEEEtran}
\usepackage{epsfig,amsmath,amssymb,epsf,cite,scalefnt,subfig,array} 
\usepackage{cases}
\usepackage{graphicx,pifont,epstopdf,grffile}

    \def\by{{\mathbf{y}}}

   \def\bS{{\mathbf{S}}}

 \def\ibb{{\pmb{b}}}  \def\ibd{{\pmb{d}}} 
  \def\ibh{{\pmb{h}}}  
    
\def\ibp{{\pmb{p}}}  \def\ibr{{\pmb{r}}}  \def\ibt{{\pmb{t}}}
\def\ibu{{\pmb{u}}}   \def\ibx{{\pmb{x}}} 
\def\ibz{{\pmb{z}}}

\def\span{\mathop{\mathrm{span}}}

\renewcommand{\baselinestretch}{1.0}

     \def\d4{\!\!\!\!}              \def\eps{\epsilon}
                       \def\al{\alpha}

 \def\bPhi{\mathbf{\Phi}}    \def\lam{\lambda} 
\def\bal{\boldsymbol{\alpha}} \def\blam{\boldsymbol{\lambda}}
\def\bzeta{\boldsymbol{\zeta}}

  \def\-{\! - \!}  \def\+{\! + \!}  \def\={\! = \!}  \def\>{\! > \!}

\newcommand{\bef}{\begin{figure}}
\newcommand{\eef}{\end{figure}}
\newcommand{\beq}{\begin{eqnarray}}
\newcommand{\eeq}{\end{eqnarray}}

\newcommand{\qed}{\nobreak \ifvmode \relax \else
\ifdim\lastskip<1.5em \hskip-\lastskip \hskip1.5em plus0em
minus0.5em \fi \nobreak \vrule height0.5em width0.5em
depth0.25em\fi}

\ifCLASSINFOpdf
\else
\fi
\begin{document}
%
\title{Robust MIMO Radar Target Localization based on Lagrange Programming Neural Network}
%
%
%
\author{Hao~Wang,
        Chi-Sing~Leung,~\IEEEmembership{Senior Member,~IEEE,}
        Hing Cheung So,~\IEEEmembership{Fellow,~IEEE,}
        Junli~Liang,~\IEEEmembership{Senior Member,~IEEE,}
        Ruibin~Feng,
        and~Zifa~Han.
\thanks{Hao~Wang, Chi-Sing~Leung, Hing Cheung So, Ruibin~Feng, and~Zifa~Han are with the Department of Electronic Engineering, City University of Hong Kong, Hong Kong.}
\thanks{Junli~Liang is with Northwestern Polytechnical University, Xi'an 710072, China.}}

%
%

\markboth{IEEE TRANSACTIONS ON ,~Vol.~1, No.~, ~2016}%
{Shell \MakeLowercase{\textit{et al.}}: Robust MIMO Radar Target Localization based on Lagrange Programming Neural Network}
%



\maketitle

\begin{abstract}
This paper focuses on target localization in a widely distributed multiple-input-multiple-output (MIMO) radar system.
In this system, range measurements, which include the sum of distances between transmitter and target and the distances from the target to receivers, are used.
We can obtain an accurate estimated position of the target by minimizing the measurement errors.
In order to make our model come closer to reality, we introduce two kinds of noises, namely, Gaussian noise and outliers.
When we evaluate a target localization algorithm, its localization accuracy and computational complexity are two main criteria.
To improve the positioning accuracy, the original problem is formulated as solving a non-smooth constrained optimization problem in which the objective function is either $l_1$-norm or $l_0$-norm term.
To achieve a real-time solution, the Lagrange programming neural network (LPNN) is utilized to solve this problem.
However, it is well known that LPNN requires twice-differentiable objective function and constraints.
Obviously, the $l_1$-norm or $l_0$-norm term in the objective function does not satisfy this requirement.
To address this non-smooth optimization problem, this paper proposes two modifications based on the LPNN framework. In the first method, a differentiable proximate $l_1$-norm function is introduced.
While in the second method, locally competitive algorithm is utilized.
Simulation and experimental results demonstrate that the performance of the proposed algorithms outperforms several existing schemes.
\end{abstract}


%
\IEEEpeerreviewmaketitle

\section{Introduction}\label{section1}

%
%
%
Generally speaking, multiple-input multiple-output (MIMO) radar systems, use multiple antennas to transmit multiple signals and employ multiple receivers to receive the echoes from the target \cite{godrich2010target}.
MIMO radar systems can be grouped into two categories, namely, colocated and distributed antennas.
The former positions its antennas closely, and utilizes the waveform diversity for performance improvement.
While for the distributed MIMO radar, its transmitters and receivers are widely separated with each other.
It employs the spatial diversity to improve its localization accuracy \cite{lpnn4}.
Compared with traditional radar systems, the MIMO radar, especially the distributed variant, has many improvements in the aspect of localization accuracy and robustness against noise.
Hence, in this paper, we focus on the target localization problem under distributed MIMO radar system.

For distributed MIMO radar system, the target localization can be obtained directly or indirectly.
For direct approaches, including the maximum likelihood (ML) estimators \cite{godrich2010target}, \cite{bar2011direct}, the position of target is directly calculated according to the measurements collected by the antennas.
These methods are based on two-dimensional (2-D) search, which requires enormous computation.
On the other hand, the indirect approaches first detect the time-of-arrival (TOA) measurements, then estimate the target position according to TOAs.
In this case, the ML methods can also be employed to estimate the target position.
Generally speaking, ML methods solve a non-convex optimization problem \cite{chalise2014target}.
Hence, the ML methods are usually transformed into the least squares (LS) method to reduce their complexity \cite{huang2001real,li2004least}.
The LS solutions are usually obtained in an iterative manner, and their accuracy is largely dependent on the initial position estimate.
Besides, due to the property of $l_2$-norm, this method is highly sensitive to outliers.
Our proposed algorithms are also based on the LS method, but we provide modifications to avoid the above mentioned disadvantages.

In this paper, we develop a robust target localization algorithms for distributed MIMO radar system based on the Lagrange programming neural network (LPNN)~\cite{lpnn0,lpnn_convergence,lpnn1,lpnn2,lpnn3,lpnn4}.
And the $l_1$-norm or $l_0$-norm term is applied as objective function to achieve robustness against outliers.
In particular, we focus on the LPNN solver to handle optimization problems with $l_1$-norm or $l_0$-norm term.
However, the LPNN framework requires that its objective function and constraints should be twice differentiable.
In the first proposed method, we introduce a differentiable proximate function to replace $l_1$-norm term in the objective function.
While, in the second method, the internal state concept of the LCA is utilized to convert the non-differentiable components due to the $l_1$-norm or $l_0$-norm as differentiable expressions.

The rest of this paper is organized as follows. Background of MIMO radar target localization, LPNN and LCA are described in Section~\ref{section2}.
In Section~\ref{section3}, two target localization algorithms are devised. The local stability of the two approaches is proved in Section~\ref{section4}.
Numerical results for algorithm evaluation and comparison are provided in Section~\ref{section5}.
Finally, conclusions are drawn in Section~\ref{section6}.
\section{Background}\label{section2}
\subsection{Notation}
We use a lower-case or upper-case letter to represent a scalar while vectors and matrices are denoted by bold lower-case and upper-case letters, respectively. The transpose operator is denoted as $(\centerdot)^ \mathrm{T}$. Other mathematical symbols are defined in their first appearance.
\subsection{MIMO Radar Localization}
A MIMO radar localization system \cite{lpnn4,liang2016robust} normally includes $M$ transmitters and $N$ receivers in a 2-D space.
The positions of these transmitters, receivers and the target to be detected are expressed as $\ibt_i=[x^t_i,y^t_i]^\mathrm{T}, i=1, \cdots ,M$, $\ibr_j=[x^r_j,y^r_j]^\mathrm{T}, j=1, \cdots ,N$ and $\ibp=[x,y]^\mathrm{T}$, respectively. Assume that each transmitter sends out a distinct electromagnetic wave.
All these electromagnetic waves are reflected by the target, and then collected by receivers.
The propagation time from the transmitter $\ibt_i$ to the target is $\tau^t_i$, while the propagation time between the target and the receiver $\ibr_j$ is $\tau^r_j$. Thus, the distance from the transmitter $\ibt_i$ to target, and that from the target to receiver $\ibr_j$ can be respectively defined as
\beq
d^t_i=\|\ibp-\ibt_i\|_2=\sqrt{(x^t_i-x)^2+(y^t_i-y)^2}, \label{eq-dt} \\
d^r_j=\|\ibp-\ibr_j\|_2=\sqrt{(x^r_j-x)^2+(y^r_j-y)^2}. \label{eq-dr}
\eeq
The total propagation distances are
\beq \label{eq-distance}
d_{i,j}=d^t_i+d^r_j, \,\,\,\,i=1, \cdots ,M, \,\,\,\,j=1, \cdots ,N.
\eeq
We see that this system needs to measure $M\times N$ distances. However, in practice, noise is almost inevitable. Therefore, the expressions of observed propagation distances are
\beq \label{eq-realdistance}
\hat{d}_{i,j}=c(\tau^t_i+\tau^r_j)=d_{i,j}+\eps_{i,j},
\eeq
where $i=1, \cdots ,M$, $j=1, \cdots ,N$, $\eps_{i,j}$ denotes the noise, $c$ is the speed of light.
The aim of this system is to estimate the position of target $\ibp$ from $\{\ibt_i\}$,$\{\ibr_j\}$ and $\{\hat{d}_{i,j}\}$.
For simplicity, most off-the-shelf algorithms directly assume that $\eps_{i,j}$ obeys a zero-mean Gaussian distribution.
While, in fact, the impulsive noise even outliers cannot be avoided in this system.
For this reason, in this paper, we consider that $\eps_{i,j}$ includes the zero-mean Gaussian white noise and some outliers.

\subsection{Lagrange Programming Neural Network}
The LPNN, introduced in \cite{lpnn0}, is an analog neural network which is able to solve the general nonlinear constrained optimization problem, given by
\begin{subequations}\label{eq-1.6}
\beq
\min\limits_{\ibz} &\,\,f(\ibz) \\
\mbox{s.t.}&\,\, \ibh(\ibz)=0,
\eeq
\end{subequations}
where $\ibz=[z_1,\cdots,z_n]^\mathrm{T}$ is the variable vector being optimized,
 $f:\mathbb{R}^n \to \mathbb{R}$ is the objective function, and $\ibh:\mathbb{R}^n \to \mathbb{R}^m$ ($m<n$) represents $m$ equality constraints. Both $f$ and $\ibh$
should be twice differentiable in LPNN framework. First, we set up the Lagrangian of \eqref{eq-1.6}:
\beq
\label{eq-1.7}
L(\ibz,\bzeta)=f(\ibz)+\bzeta^\mathrm{T}\ibh(\ibz),
\eeq
where $\bzeta=[\zeta_1,\cdots,\zeta_m]^\mathrm{T}$ is the Lagrange multiplier vector. In LPNN framework, there are $n$ variable neurons and $m$ Lagrangian neurons, which are used to hold the state variable vector $\ibz$ and the Lagrange multiplier vector $\bzeta$, respectively.
The dynamics of the neurons can be expressed as
\begin{subequations}\label{eq-1.8}
\beq
\frac{d\ibz}{dt}=& \displaystyle -\frac{\partial L(\ibz,\bzeta)}{\partial \ibz} \\
\frac{d\bzeta}{dt}=& \displaystyle \frac{\partial L(\ibz,\bzeta)}{\partial \bzeta}.
\eeq
\end{subequations}
After the neurons settle down at an equilibrium point, the output of the neurons is the solution we want.
The purpose of dynamic in (\ref{eq-1.8}a) is to seek for a state with the minimum objective value, while (\ref{eq-1.8}b) aims to constrain its outputs into the feasible region.
The network will settle down at a stable state if several conditions are satisfied~\cite{lpnn0,lpnn3,lpnn4}.
Obviously, $f$ and $\ibh$ should be differentiable. Otherwise, the dynamics in (\ref{eq-1.8})  cannot be defined.
\subsection{Locally Competitive Algorithm}
The LCA \cite{lca0} is also an analog neural network which is designed for solving the following unconstrained optimization problem:
\beq\label{eq-1.9}
L_{\rm lca}= \frac{1}{2} \| \ibb -\bPhi \ibz \|_2^2  + \lambda \| \ibz \|_1
\eeq
where $\ibz \in \mathbb{R}^n$, $\ibb \in \mathbb{R}^m$, $\bPhi\in \mathbb{R}^{m\times n} (m < n)$.
To construct the dynamics of LCA, we need to calculate the gradient of $L_{\rm lca}$ with respect to $\ibz$. However, $\lambda \| \ibz \|_1$ is non-differentiable at zero point. Thus the gradient of \eqref{eq-1.9} is
\beq\label{eq-1.10}
\partial_{\ibz} L_{\rm lca}=  -\bPhi(\ibb -\bPhi \ibz)+\lambda \partial \| \ibz \|_1,
\eeq
where $\partial \| \ibz \|_1$ denotes the sub-differential of $\| \ibz \|_1$. According to the definition of sub-differential, we know that at the non-differentiable point the sub-differential is equal to a set\footnote{For the absolute function $|z|$, the sub-differential $\partial |z|$ at $z=0$ is equal to $[-1,1]$.}.
To handle this issue, LCA introduces an internal state vector $\ibu=[u_1,\cdots,u_n]^\mathrm{T}$ and defines a relationship between $\ibu$ and $\ibz$,
\begin{equation}
z_i = T_{\lambda}(u_i) = \left\{ \begin{array}{lcl}
0, &  |u_i| \leq \lambda, \\
u_i - \lambda \mbox{sign} (u_i),  & |u_i| > \lambda.
\end{array}\right.
\label{internal2}
\end{equation}
In the LCA, $\ibz$ and  $\ibu$ are known as the output state variable and internal state variable vectors, respectively, $\lambda$ denotes the threshold of the function.
Furthermore, from~\eqref{internal2}, we can deduce that
\begin{equation}
\label{internal1x}
\ibu - \ibz \in \lambda \partial \|\ibz \|_1.
\end{equation}
Hence, LCA defines its dynamics on $\ibu$ rather than $\ibz$ as
\begin{equation}
\frac{d \ibu}{dt}=-\partial_{ \ibz } L_{\rm lca}= - \ibu + \ibz + \bPhi^\mathrm{T}(\ibb - \bPhi \ibz).
\label{eqn:dyna}
\end{equation}
It is worth noting that if the dynamics of $\ibz$ are used directly, we need to calculate $\partial \|\ibz \|_1$ which
is equal to a set at the zero point. Therefore, LCA uses ${d \ibu}/{dt}$ rather than ${d \ibz}/{dt}$.

In~\cite{lca0}, a general element-wise threshold function is also proposed, which is described as
\beq\label{eq-1.14}
z_i=T_{(\eta,\delta,\lambda)}(u_i)=\mbox{sign} (u_i)\frac{ |u_i|-\delta\lambda}{1 + e^{-\eta(|u_i|-\lambda)}},
\eeq
where $\lambda$ still denotes the threshold, $\eta$ is a parameter used to control the threshold transition rate, and $\delta \in [0, 1]$  indicates adjustment fraction after the internal neuron across threshold \cite{lca0}.
Some examples of this general threshold function are given in Fig.1.
The general threshold function in \eqref{eq-1.14} is used for solving the unconstrained optimization problem given by
\beq\label{eq-general}
\tilde{L}_{\rm lca}=  \frac{1}{2} \| \ibb -\bPhi \ibz \|_2^2 + \lambda\bS_{\eta,\delta,\lambda}(\ibz).
\eeq
where $\bS_{\eta,\delta,\lambda}(\ibz)$ is a proximate function of $L_p$-norm $(0\leq p\leq1)$, and it has an important property:
\begin{equation}
\label{internal1x2016}
\lambda \frac{\partial \bS_{\eta,\delta,\lambda}(\ibz)}{\partial \ibz} \equiv \ibu - \ibz.
\end{equation}
The exact form of $\bS_{\eta,\delta,\lambda}(\ibz)$ cannot be obtained.
However, it does not influence the application of it due to the fact that the neural dynamics are defined in terms of its gradient function rather than the penalty term itself. According to the discussion in~\cite{lca0}, when $\eta \rightarrow \infty$ and $\delta=0$, we obtain an ideal hard threshold function given by
\beq
\label{eq-thresholdl0a}
z_i=T_{(\infty,0,\lambda)}(u_i)=
\left\{ \begin{array}{lcl}
0,  & |u_i| \leq \lambda, \\
u_i, & |u_i| > \lambda.
\end{array}\right.
\eeq
The corresponding penalty term is
\beq\label{eq-generalL0}
\lambda\bS_{\infty,0,\lambda}(\ibz) = \frac{1}{2} \sum_{i=1}^n \mathcal{I}(|z_i|>\lambda),
\eeq
where $\mathcal{I}(\cdot)$ denotes an indicator function.
Obviously, $\bS_{\infty,0,\lambda}(\ibz)$ is a proximate function of $l_0$-norm.
It is worth noting that the variables $z_i$ produced by the ideal threshold function (\ref{eq-thresholdl0a}) cannot take values in the range of $[-1,0)$ and $(0,1]$.

While, if we let $\eta \rightarrow \infty$ and $\delta=1$, then the general threshold function is reduced to the soft threshold function, given by
\beq
\label{eq-thresholdl0aa}
z_i=T_{(\infty,1,\lambda)}(u_i)=T_{\lambda}(u_i)
\eeq
and the penalty term
\beq\label{eq-generalL1aa}
\lambda\bS_{\infty,1,\lambda}(\ibz) = \lambda \|\ibz\|_1.
\eeq
More details about parameter setting of the threshold function can be found in \cite{lca0}. Besides, the behavior of the dynamics under different settings has been studied in~\cite{lca0,lca_convergence1,lca_convergence2}.

\begin{figure}[htb]
\centering
\centerline{\includegraphics[width=3.5in]{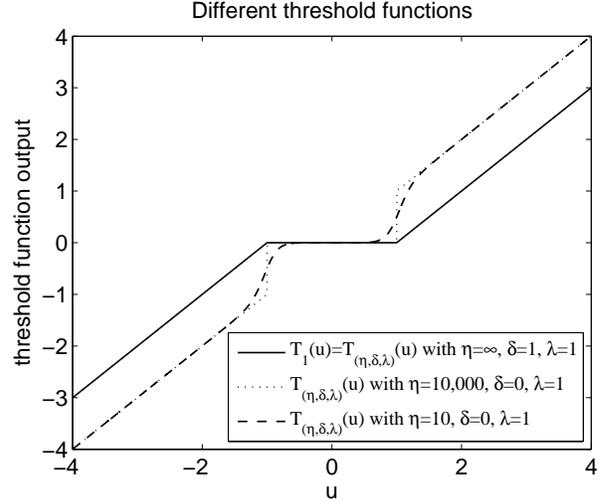}}
\caption{Examples of general threshold function.}
\label{threshold}
\end{figure}

\section{Development of Proposed Method} \label{section3}
\subsection{Problem Formulation}
In traditional TOA system, it is generally assumed that the noise $\eps_{i,j}(i=1, \cdots ,M, j=1, \cdots ,N)$ follows Gaussian distribution. And according to LS method \cite{chalise2014target,lpnn4}, the problem can be formulated as
\beq
\min\limits_{\ibp} \,\,\sum_{i=1}^M\sum_{j=1}^N \left(\hat{d}_{i,j}-\|\ibp-\ibt_i\|_2-\|\ibp-\ibr_j\|_2\right)^2.
\eeq
Combining with the definitions given in \eqref{eq-dt} and \eqref{eq-dr}, the problem can be rewritten as
\begin{subequations}
\label{eq-formulation_l2_0}
\beq
\min\limits_{\ibp,d^t_i,d^r_j} \,\, & \sum_{i=1}^M\sum_{j=1}^N \left(\hat{d}_{i,j}-d^t_i-d^r_j\right)^2,\\
\mbox{s.t.}\,\, &  d^t_i=\|\ibp-\ibt_i\|_2, \,\,\,\, i=1, \cdots ,M,\\
&  d^r_j=\|\ibp-\ibr_j\|_2, \,\,\,\, j=1, \cdots ,N.
\eeq
\end{subequations}
Denote,
\beq
&\ibd^t=[d^t_1,...,d^t_M,d^t_1,...,d^t_M,...,d^t_1,...,d^t_M]^\mathrm{T}, \nonumber \\
&\ibd^r=[d^r_1,...,d^r_1,d^r_2,...,d^r_2,...,d^r_N,...,d^r_N]^\mathrm{T}, \nonumber \\
&\hat{\ibd}=[\hat{d}_{1,1},...,\hat{d}_{M,1},\hat{d}_{1,2},...,\hat{d}_{M,2}, ... ,\hat{d}_{1,N},...,\hat{d}_{M,N},]^\mathrm{T}, \nonumber
\eeq
where they are all $MN\times 1$ vectors.
Then \eqref{eq-formulation_l2_0} can be modified as
\begin{subequations}
\label{eq-formulation_l2_1}
\beq
\min\limits_{\ibp,d^t_i,d^r_j} \,\, & \left\| \hat{\ibd}-\ibd^t-\ibd^r \right\|_2^2,\\
\mbox{s.t.}\,\, &  {d^t_i}^2=\|\ibp-\ibt_i\|_2^2, \,\,\,\, i=1, \cdots ,M,\\
&  {d^r_j}^2=\|\ibp-\ibr_j\|_2^2, \,\,\,\, j=1, \cdots ,N,\\
&  d^t_i\geq0,\,\,\,\, i=1, \cdots ,M\\
&  d^t_i\geq0,\,\,\,\, j=1, \cdots ,N.
\eeq
\end{subequations}
It is well known that, compared with $l_2$-norm, $l_1$-norm is less sensitive to outliers, hence in our proposed model, the problem is modified as the following form:
\begin{subequations}
\label{eq-formulation_l1}
\beq
\min\limits_{\ibp,d^t_i,d^r_j} \,\, & \left\| \hat{\ibd}-\ibd^t-\ibd^r \right\|_1,\\
\mbox{s.t.}\,\, &  {d^t_i}^2=\|\ibp-\ibt_i\|_2^2, \,\,\,\, i=1, \cdots ,M,\\
&  {d^r_j}^2=\|\ibp-\ibr_j\|_2^2, \,\,\,\, j=1, \cdots ,N,\\
&  d^t_i\geq0,\,\,\,\, i=1, \cdots ,M, \\
&  d^t_i\geq0,\,\,\,\, j=1, \cdots ,N.
\eeq
\end{subequations}
\eqref{eq-formulation_l1} includes $M+N$ equality constraints and $M+N$ inequality constraints.
Since LPNN can handle the problem with equality constraint only, the inequality constraints in \eqref{eq-formulation_l1} should be removed.
To achieve this purpose, we use the following proposition.

\textbf{Proposition 1}: The optimization problem in \eqref{eq-formulation_l1} is equivalent to
\begin{subequations}
\label{eq-formulation_l1_1}
\beq
\min\limits_{\ibp,d^t_i,d^r_j} \,\, & \left\| \hat{\ibd}-\ibd^t-\ibd^r \right\|_1,\\
\mbox{s.t.}\,\, &  {d^t_i}^2=\|\ibp-\ibt_i\|_2^2, \,\,\,\, i=1, \cdots ,M,\\
&  {d^r_j}^2=\|\ibp-\ibr_j\|_2^2, \,\,\,\, j=1, \cdots ,N.
\eeq
\end{subequations}

Proof: Obviously, to simplify the optimization problem \eqref{eq-formulation_l1} to \eqref{eq-formulation_l1_1}, we need to prove that the inequality constraints in \eqref{eq-formulation_l1} are unnecessary.
Suppose that $(\ibp^*,{d^t_1}^*,...,{d^t_M}^*,{d^r_1}^*,...,{d^r_N}^*)$ is the optimal solution of \eqref{eq-formulation_l1_1}.
According to the reverse triangle inequality, we see that
\beq
\label{eq:triangle_inequality0}
\sum_{i=1}^M\sum_{j=1}^N \left|\hat{d}_{i,j}-{d^t_i}^*-{d^r_j}^*\right| \geq \sum_{i=1}^M\sum_{j=1}^N|\hat{d}_{i,j}|-|{d^t_i}^*|-|{d^r_j}^*|,
\eeq
As $\hat{d}_{i,j}$ is distance, all of them must be nonnegative. Thus,  \eqref{eq:triangle_inequality0} can be rewritten as
\beq
\label{eq:triangle_inequality1}
\sum_{i=1}^M\sum_{j=1}^N \left|\hat{d}_{i,j}-{d^t_i}^*-{d^r_j}^*\right| \geq \sum_{i=1}^M\sum_{j=1}^N\hat{d}_{i,j}-|{d^t_i}^*|-|{d^r_j}^*|
\eeq
The inequality in \eqref{eq:triangle_inequality1} implies that the optimal solution $(\ibp^*,{d^t_1}^*,...,{d^t_M}^*,{d^r_1}^*,...,{d^r_N}^*)$ of \eqref{eq-formulation_l1} is greater or equal to the solution achieved by the feasible point $(\ibp^*,|{d^t_1}^*|,...,|{d^t_M}^*|,|{d^r_1}^*|,...,|{d^r_N}^*|)$. Since $(\ibp^*,{d^t_1}^*,...,{d^t_M}^*,{d^r_1}^*,...,{d^r_N}^*)$ is the optimal solution, the equality of \eqref{eq:triangle_inequality1} must be held. Thus ${d^t_i}^*=|{d^t_i}^*|$ for $\forall i \in [1, \cdots ,M]$, and ${d^r_j}^*=|{d^r_j}^*|$ for $j \in [1, \cdots ,N]$.
In other words, ${d^t_i}^*\geq 0$ for $i \in [1, \cdots ,M]$, and ${d^r_j}^*\geq 0$ for $\forall j \in [1, \cdots ,N]$. Hence, we can remove the inequality constraints in \eqref{eq-formulation_l1}. $\blacksquare$

In order to facilitate the analysis of this paper, we introduce a dummy variable $\ibz$ and rewrite \eqref{eq-formulation_l1_1} as
\begin{subequations}
\label{eq-formulation_l1_2}
\beq
\min\limits_{\ibp,\ibz,d^t_i,d^r_j} \,\, & \|\ibz\|_1,\\
\mbox{s.t.}\,\,  &\ibz=\hat{\ibd}-\ibd^t-\ibd^r, \\
&  {d^t_i}^2=\|\ibp-\ibt_i\|_2^2, \,\,\,\, i=1, \cdots ,M,\\
&  {d^r_j}^2=\|\ibp-\ibr_j\|_2^2, \,\,\,\, j=1, \cdots ,N,
\eeq
\end{subequations}
where $\ibz=[z_{1,1},...,z_{M,1},z_{1,2},...,z_{M,2}, ... ,z_{1,N},...,z_{M,N}]^\mathrm{T}, \nonumber$ is a vector with $M\times N$ elements.
\subsection{LPNN for MIMO Radar Localization}
To obtain a real-time solution of the problem given in \eqref{eq-formulation_l1_2}, we consider using the LPNN framework. However, LPNN requires its objective function and constraints are all twice differentiable. Obviously, due to the $l_1$-norm term, the objective function in \eqref{eq-formulation_l1_2} does not satisfy this requirement. Hence, for calculating the gradient of $l_1$-norm at the non-differentiable point, we propose two approaches.

\begin{figure}[!ht]
\centering
\centerline{\includegraphics[width=3.2in]{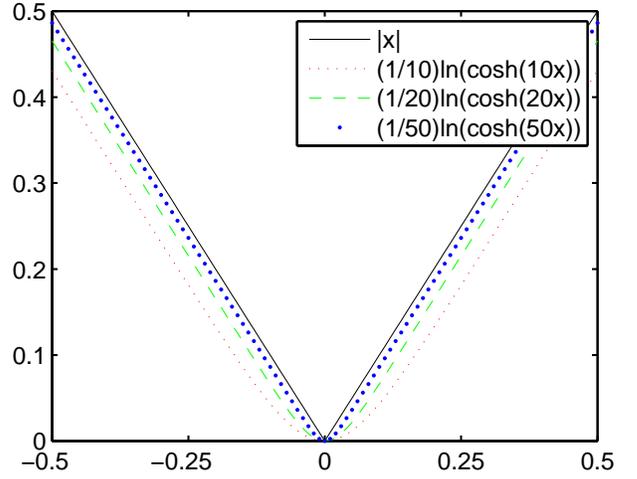}}
\caption{$(1/a)\ln(\cosh(ax_i))$.}
\label{fig-proximatel1}
\end{figure}

\textbf{Method 1}:
Intuitively, we consider using a differentiable $l_1$-norm proximate function \cite{leung2014recurrent}:
\beq\label{eq-proximatel1}
g(x)=\frac{\ln(\cosh(a x))}{a},
\eeq
where $a>1$ is a scalar. Fig.\ref{fig-proximatel1} shows the shapes of $g(x)$ with different $a$. It is observed that the shape of the proximate function is quite similar with $l_1$-norm when $a$ is large.

With this $l_1$-norm proximate function, the problem in \eqref{eq-formulation_l1_2} can be rewritten as
\begin{subequations}
\label{eq-method1_0}
\beq
\min\limits_{\ibp,z_k,d^t_i,d^r_j} \,\, & \sum_{k=1}^{MN} \frac{1}{a}\ln(\cosh(a z_k)),\\
\mbox{s.t.}\,\,  &\ibz=\hat{\ibd}-\ibd^t-\ibd^r, \\
&  {d^t_i}^2=\|\ibp-\ibt_i\|_2^2, \,\,\,\, i=1, \cdots ,M,\\
&  {d^r_j}^2=\|\ibp-\ibr_j\|_2^2, \,\,\,\, j=1, \cdots ,N.
\eeq
\end{subequations}
After the modification, the objective function is differentiable.
It is worth noting that the gradient of $(1/a)\sum_{k=1}^{MN}\ln(\cosh(a z_k)$ with respect to $z_k$ is equal to $\tanh(a z_k)$ (the hyperbolic tangent function) which is frequently used as an activation function in artificial neural networks.

Theoretically, we can directly use \eqref{eq-method1_0} to deduce the LPNN dynamics. While, our preliminary simulation results show that the neural dynamics may not be stable. Hence, we introduce several augmented terms into its objective (\ref{eq-method1_0}a), after that we have
\begin{subequations}
\label{eq-method1_1}
\beq
\min\limits_{\ibp,\by,d^t_i,d^r_j} \,\, & \sum_{k=1}^{MN} \frac{1}{a}\ln(\cosh(az_k)) \nonumber\\
&+\frac{C}{2}\|\ibz-\hat{\ibd}+\ibd^t+\ibd^r\|_2^2 \nonumber \\
&+\frac{C}{2}\sum_{i=1}^M\left({d^t_i}^2-\|\ibp-\ibt_i\|_2^2\right)^2\nonumber \\ &+\frac{C}{2}\sum_{j=1}^N\left({d^r_j}^2-\|\ibp-\ibr_j\|_2^2\right)^2 \\
\mbox{s.t.}\,\,  &\ibz=\hat{\ibd}-\ibd^t-\ibd^r, \\
&  {d^t_i}^2=\|\ibp-\ibt_i\|_2^2, \,\,\,\, i=1, \cdots ,M,\\
&  {d^r_j}^2=\|\ibp-\ibr_j\|_2^2, \,\,\,\, j=1, \cdots ,N,
\eeq
\end{subequations}
where $C$ is a scalar used for regulating the magnitude of these augmented terms.
For any point in feasible region, the constraints must be satisfied.
Hence the augmented terms equal $0$ at an equilibrium point.
They can improve the stability and convexity of the dynamics and do not influence the optimal solution.
The Lagrangian of \eqref{eq-method1_1} is
\beq\label{eq-method1Lagrangian1}
L(\ibp,\ibz,d^t_i,d^r_j,\bal,\boldsymbol{\beta},\boldsymbol{\lam})=\sum_{k=1}^{MN} \frac{1}{a}\ln(\cosh(az_k)) \nonumber \\
+\bal^\mathrm{T}(\ibz-\hat{\ibd}+\ibd^t+\ibd^r) \nonumber \\
+\sum_{i=1}^M\beta_i({d^t_i}^2-\|\ibp-\ibt_i\|_2^2)\nonumber \\ +\sum_{j=1}^N\lam_j({d^r_j}^2-\|\ibp-\ibr_j\|_2^2)\nonumber \\
+\frac{C}{2}\|\ibz-\hat{\ibd}+\ibd^t+\ibd^r\|_2^2 \nonumber \\
+\frac{C}{2}\sum_{i=1}^M\left({d^t_i}^2-\|\ibp-\ibt_i\|_2^2\right)^2\nonumber \\
+\frac{C}{2}\sum_{j=1}^N\left({d^r_j}^2-\|\ibp-\ibr_j\|_2^2\right)^2
\eeq
where $\ibp, \ibz, d^t_i(i=1, \cdots ,M), d^r_j(j=1, \cdots ,N)$ are state variable vectors, $\bal=[\al_1,...,\al_{MN}]^\mathrm{T}$, $\boldsymbol{\beta}=[\beta_1,...,\beta_M]^\mathrm{T}$ and $\boldsymbol{\lam}=[\lam_1,...,\lam_N]^\mathrm{T}$ are Lagrangian variable vectors.
According to this Lagrangian and the concepts of LPNN given in \eqref{eq-1.8}, its dynamics are defined as:
\beq
\frac{d\ibz}{dt}&=&
-\frac{\partial L(\ibp,\ibz,d^t_i,d^r_j,\bal,\boldsymbol{\beta},\boldsymbol{\lam})}{\partial \ibz} \nonumber \\ &=&-\tanh (a\ibz)-\bal-C(\ibz-\hat{\ibd}+\ibd^t+\ibd^r), \label{method1dynamic1_1} \\
\frac{dd^t_i}{dt}&=&
-\frac{\partial L(\ibp,\ibz,d^t_i,d^r_j,\bal,\boldsymbol{\beta},\boldsymbol{\lam})}{\partial d^t_i} \nonumber \\
&=&-{\sum_{j=1}^{N-1}\al_{i+jM}}-2\beta_id^t_i\nonumber \\ &&-C{\sum_{j=1}^{N}(z_{i,j}-\hat{d}_{i,j}+d^t_i+d^r_j)} \nonumber \\ &&-2Cd^t_i({d^t_i}^2-\|\ibp-\ibt_i\|_2^2), \label{method1dynamic1_2} \\
\frac{dd^r_j}{dt}&=&
-\frac{\partial L(\ibp,\ibz,d^t_i,d^r_j,\bal,\boldsymbol{\beta},\boldsymbol{\lam})}{\partial d^r_j} \nonumber\\
&=&-{\sum_{i=1}^{M}\al_{i+(j-1)M}}-2\lam_jd^r_j\nonumber \\ &&-C{\sum_{i=1}^{M}(z_{i,j}-\hat{d}_{i,j}+d^t_i+d^r_j)} \nonumber \\ &&-2Cd^r_j({d^r_j}^2-\|\ibp-\ibr_j\|_2^2), \label{method1dynamic1_3}\\
\frac{d\ibp}{dt}&=&
-\frac{\partial L(\ibp,\ibz,d^t_i,d^r_j,\bal,\boldsymbol{\beta},\boldsymbol{\lam})}{\partial \ibp} \nonumber\\
&=&\sum_{i=1}^{M}2\beta_i(\ibp-\ibt_i)+\sum_{j=1}^{N}2\lam_j(\ibp-\ibr_j)\nonumber \\ &&+2C\sum_{i=1}^{M}(\ibp-\ibt_i)({d^t_i}^2-\|\ibp-\ibt_i\|_2^2) \nonumber \\ &&+2C\sum_{j=1}^{N}(\ibp-\ibr_j)({d^r_j}^2-\|\ibp-\ibr_j\|_2^2), \label{method1dynamic1_4}
\eeq
\beq
\frac{d\bal}{dt}&=&
-\frac{\partial L(\ibp,\ibz,d^t_i,d^r_j,\bal,\boldsymbol{\beta},\boldsymbol{\lam})}{\partial \bal} \nonumber\\ &=&\ibz-\hat{\ibd}+\ibd^t+\ibd^r, \label{method1dynamic1_5} \\
\frac{d\beta_i}{dt}&=&
-\frac{\partial L(\ibp,\ibz,d^t_i,d^r_j,\bal,\boldsymbol{\beta},\boldsymbol{\lam})}{\partial \beta_i} \nonumber\\
&=&{d^t_i}^2-\|\ibp-\ibt_i\|_2^2, \label{method1dynamic1_6} \\
\frac{d\lam_j}{dt}&=&
-\frac{\partial L(\ibp,\ibz,d^t_i,d^r_j,\bal,\boldsymbol{\beta},\boldsymbol{\lam})}{\partial \lam_j} \nonumber\\
&=&{d^r_j}^2-\|\ibp-\ibr_j\|_2^2. \label{method1dynamic1_7}
\eeq
In the above dynamics, $i=[1,\cdots,M]$ and $j=[1,\cdots,N]$. Equations \eqref{method1dynamic1_1}-\eqref{method1dynamic1_4} are used for seeking the minimum objective value, while \eqref{method1dynamic1_5}-\eqref{method1dynamic1_7} can restrict the equilibrium point into the feasible region.

\textbf{Method 2}:
In this method, we introduce LCA into LPNN framework to solve sub-differentiable problem in \eqref{eq-formulation_l1_2}. First, we also introduce several augmented terms into the objective function to make the system more stable. Thus, \eqref{eq-formulation_l1_2} can be modified as:
\begin{subequations}
\label{eq-method2_l1}
\beq
\min\limits_{\ibp,\ibz,d^t_i,d^r_j} \,\, & \|\ibz\|_1+\frac{C}{2}\|\ibz-\hat{\ibd}+\ibd^t+\ibd^r\|_2^2 \nonumber \\ &+\frac{C}{2}\sum_{i=1}^M\left({d^t_i}^2-\|\ibp-\ibt_i\|_2^2\right)^2\nonumber \\ &+\frac{C}{2}\sum_{j=1}^N\left({d^r_j}^2-\|\ibp-\ibr_j\|_2^2\right)^2, \\
\mbox{s.t.}\,\,  &\ibz=\hat{\ibd}-\ibd^t-\ibd^r, \\
&  {d^t_i}^2=\|\ibp-\ibt_i\|_2^2, \,\,\,\, i=1, \cdots ,M,\\
&  {d^r_j}^2=\|\ibp-\ibr_j\|_2^2, \,\,\,\, j=1, \cdots ,N.
\eeq
\end{subequations}
Its Lagrangian is given by
\beq\label{eq-method2Lagrangian0}
L(\ibp,\ibz,d^t_i,d^r_j,\bal,\boldsymbol{\beta},\boldsymbol{\lam})= \|\ibz\|_1 +\bal^\mathrm{T}(\ibz-\hat{\ibd}+\ibd^t+\ibd^r) \nonumber \\
+\sum_{i=1}^M\beta_i({d^t_i}^2-\|\ibp-\ibt_i\|_2^2)\nonumber \\ +\sum_{j=1}^N\lam_j({d^r_j}^2-\|\ibp-\ibr_j\|_2^2)\nonumber \\
+\frac{C}{2}\|\ibz-\hat{\ibd}+\ibd^t+\ibd^r\|_2^2 \nonumber \\
+\frac{C}{2}\sum_{i=1}^M\left({d^t_i}^2-\|\ibp-\ibt_i\|_2^2\right)^2\nonumber \\
+\frac{C}{2}\sum_{j=1}^N\left({d^r_j}^2-\|\ibp-\ibr_j\|_2^2\right)^2
\eeq

According the concept of LCA, we introduce an internal variables $\ibu=[u_1, ..., u_{MN}]^\mathrm{T}$. Combining the dynamics of LPNN in (\ref{eq-1.8}) with the concept of LCA in \eqref{eqn:dyna}, we can deduce that:
\beq
\frac{d\ibu}{dt}&=&
-\frac{\partial L(\ibp,\ibz,d^t_i,d^r_j,\bal,\boldsymbol{\beta},\boldsymbol{\lam})}{\partial \ibz}\nonumber \\
&=&-\ibu+\ibz-\bal-C(\ibz-\hat{\ibd}+\ibd^t+\ibd^r).
\label{method2dynamic1_1} \\
\frac{dd^t_i}{dt}&=&
-\frac{\partial L(\ibp,\ibz,d^t_i,d^r_j,\bal,\boldsymbol{\beta},\boldsymbol{\lam})}{\partial d^t_i} \nonumber \\
&=&-{\sum_{j=1}^{N-1}\al_{i+jM}}-2\beta_id^t_i\nonumber \\ &&-C{\sum_{j=1}^{N}(z_{i,j}-\hat{d}_{i,j}+d^t_i+d^r_j)}\nonumber \\ &&-2Cd^t_i({d^t_i}^2-\|\ibp-\ibt_i\|_2^2), \label{method2dynamic1_2}
\eeq
\beq
\frac{dd^r_j}{dt}&=&
-\frac{\partial L(\ibp,\ibz,d^t_i,d^r_j,\bal,\boldsymbol{\beta},\boldsymbol{\lam})}{\partial d^r_j} \nonumber\\
&=&-{\sum_{i=1}^{M}\al_{i+(j-1)M}}-2\lam_jd^r_j\nonumber \\ &&-C{\sum_{j=1}^{N}(z_{i,j}-\hat{d}_{i,j}+d^t_i+d^r_j)}\nonumber \\ &&-2Cd^r_j({d^r_j}^2-\|\ibp-\ibr_j\|_2^2), \label{method2dynamic1_3} \\
\frac{d\ibp}{dt}&=&
-\frac{\partial L(\ibp,\ibz,d^t_i,d^r_j,\bal,\boldsymbol{\beta},\boldsymbol{\lam})}{\partial \ibp} \nonumber\\
&=&\sum_{i=1}^{M}2\beta_i(\ibp-\ibt_i)+\sum_{j=1}^{N}2\lam_j(\ibp-\ibr_j)\nonumber \\ &&+2C\sum_{i=1}^{M}(\ibp-\ibt_i)({d^t_i}^2-\|\ibp-\ibt_i\|_2^2) \nonumber \\ &&+2C\sum_{j=1}^{N}(\ibp-\ibr_j)({d^r_j}^2-\|\ibp-\ibr_j\|_2^2), \label{method2dynamic1_4} \\
\frac{d\bal}{dt}&=&
-\frac{\partial L(\ibp,\ibz,d^t_i,d^r_j,\bal,\boldsymbol{\beta},\boldsymbol{\lam})}{\partial \bal} \nonumber\\ &=&\ibz-\hat{\ibd}+\ibd^t+\ibd^r, \label{method2dynamic1_5} \\
\frac{d\beta_i}{dt}&=&
-\frac{\partial L(\ibp,\ibz,d^t_i,d^r_j,\bal,\boldsymbol{\beta},\boldsymbol{\lam})}{\partial \beta_i} \nonumber\\
&=&{d^t_i}^2-\|\ibp-\ibt_i\|_2^2, \label{method2dynamic1_6} \\
\frac{d\lam_j}{dt}&=&
-\frac{\partial L(\ibp,\ibz,d^t_i,d^r_j,\bal,\boldsymbol{\beta},\boldsymbol{\lam})}{\partial \lam_j} \nonumber\\
&=&{d^r_j}^2-\|\ibp-\ibr_j\|_2^2. \label{method2dynamic1_7}
\eeq
In above dynamics, $i=[1,\cdots,M]$ and $j=[1,\cdots,N]$. It is worth noting that the relationship between $\ibu$ and $\ibz$ is given by \eqref{eq-1.14}. When we set $\eta=\infty$, $\delta=1$ and $\lambda=1$, the mapping given by \eqref{eq-1.14} becomes the soft threshold and $\ibu - \ibz\in \partial \|\ibz \|_1$.
However, in order to further reduce the influence of outliers, we replace the $l_1$-norm term in the objective function of \eqref{eq-method2_l1} by $\bS_{10000,0,1}(\ibz)$, thus the function can be modified as
\begin{subequations}
\label{eq-method2_l0}
\beq
\min\limits_{\ibp,\ibz,d^t_i,d^r_j} \,\, & \bS_{10000,0,1}(\ibz)+\frac{C}{2}\|\ibz-\hat{\ibd}+\ibd^t+\ibd^r\|_2^2 \nonumber \\ &+\frac{C}{2}\sum_{i=1}^M\left({d^t_i}^2-\|\ibp-\ibt_i\|_2^2\right)^2\nonumber \\ &+\frac{C}{2}\sum_{j=1}^N\left({d^r_j}^2-\|\ibp-\ibr_j\|_2^2\right)^2, \\
\mbox{s.t.}\,\,  &\ibz=\hat{\ibd}-\ibd^t-\ibd^r, \\
&  {d^t_i}^2=\|\ibp-\ibt_i\|_2^2 \,\,\,\, i=1, \cdots ,M,\\
&  {d^r_j}^2=\|\ibp-\ibr_j\|_2^2\,\,\,\, j=1, \cdots ,N.
\eeq
\end{subequations}
For this problem we can also solve it with LPNN and LCA. The dynamics are same as \eqref{method2dynamic1_1}-\eqref{method2dynamic1_7}. The relationship between $\ibu$ and $\ibz$ is still given by \eqref{eq-1.14}.
While in this case, we set $\eta=10000$, $\delta=0$ and $\lambda=1$.
The threshold function given in \eqref{eq-1.14} becomes a proximate hard threshold and $\partial \bS_{10000,0,1}(\ibz) = \ibu - \ibz$.

For both two methods mentioned above, the dynamics are updated with the following rule,
\begin{eqnarray}
\ibu^{(l+1)} & = & \ibu^{(l)} + \mu_1\frac{d\ibu^{(l)}}{dt}, \\
{d^t_i}^{(l+1)} & = & {d^t_i}^{(l)}+\mu_2\frac{{dd^t_i}^{(l)}}{dt}, \,\,\,\, i=1, \cdots, M,\\
{d^r_j}^{(l+1)} & = & {d^r_j}^{(l)}+\mu_3\frac{{dd^r_j}^{(l)}}{dt}, \,\,\,\, j=1, \cdots, N,\\
\ibp^{(l+1)}  & = & \ibp^{(l)} +\mu_4\frac{d\ibp^{(l)} }{dt}, \\
\bal^{(l+1)}  & = &\bal^{(l)} +\mu_5\frac{d\bal^{(l)} }{dt},\\
\beta_i^{(l+1)}  & = & \beta_i^{(l)} +\mu_6\frac{d\beta_i^{(l)} }{dt},\,\,\,\, i=1, \cdots, M,\\
\lam_j^{(l+1)}  & = & \lam_j^{(l)} +\mu_7\frac{d\lam_j^{(l)} }{dt}, \,\,\,\, j=1, \cdots, N.
\end{eqnarray}
where $^{(l)}$ corresponds to the estimate at the $l$th iteration and $\mu_1,\mu_2,\mu_3,\mu_4,\mu_5,\mu_6,\mu_7$ are the step sizes which should be positive and not too large to avoid divergence.

\begin{figure}[!htb]
\centering
\subfloat[]{
\includegraphics[width=1.5in]{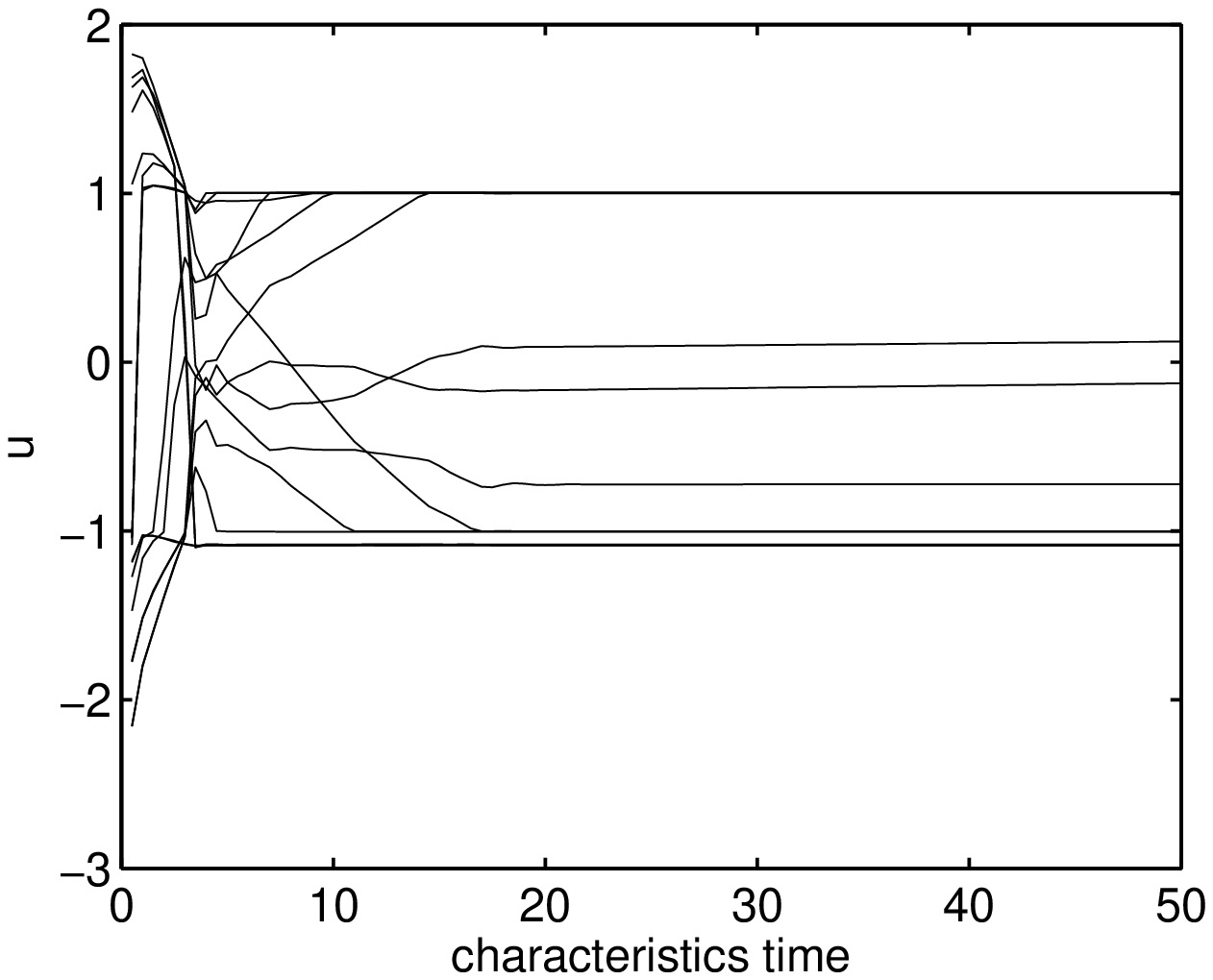}}
\subfloat[]{
\includegraphics[width=1.5in]{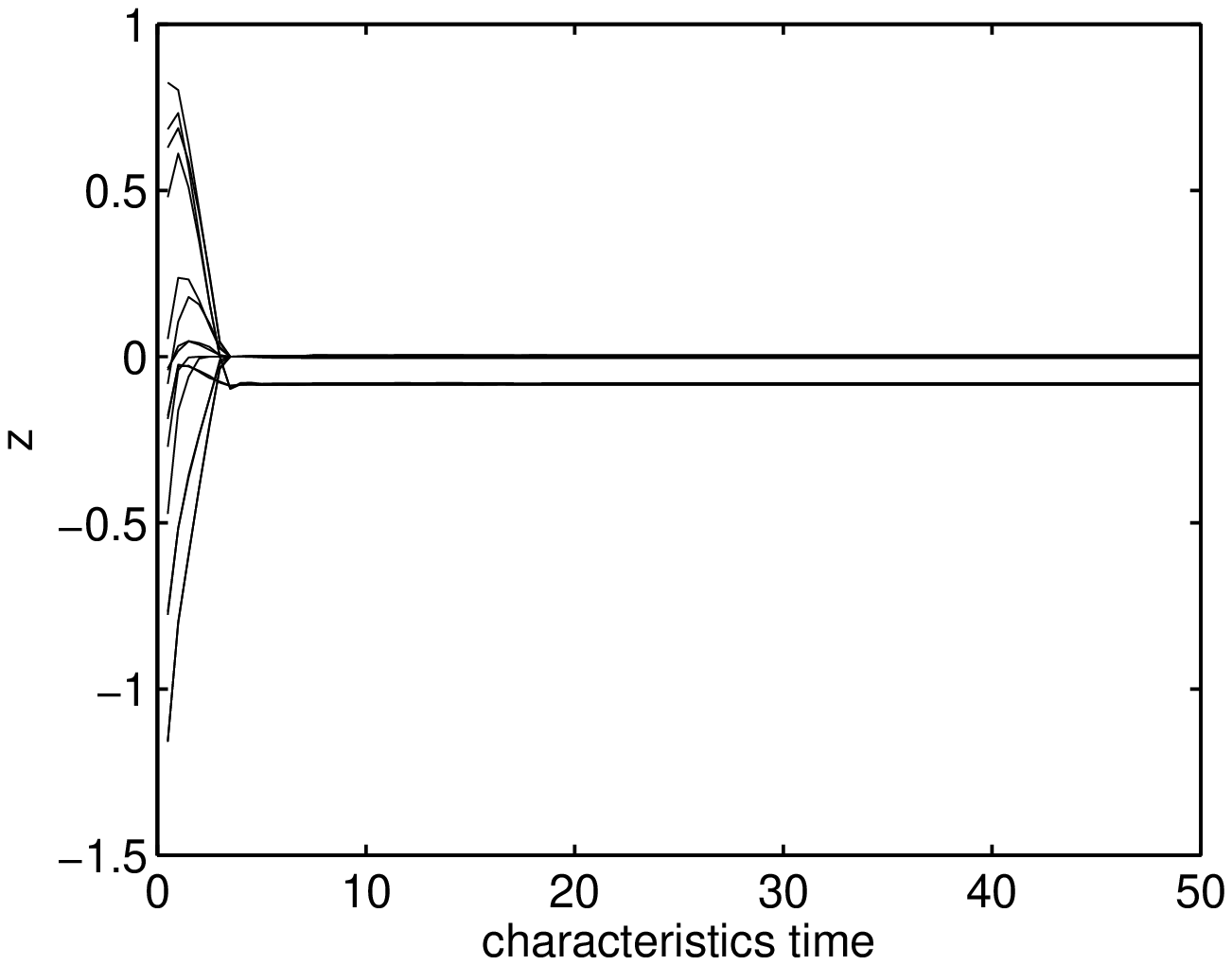}}
\hfil
\subfloat[]{
\includegraphics[width=1.5in]{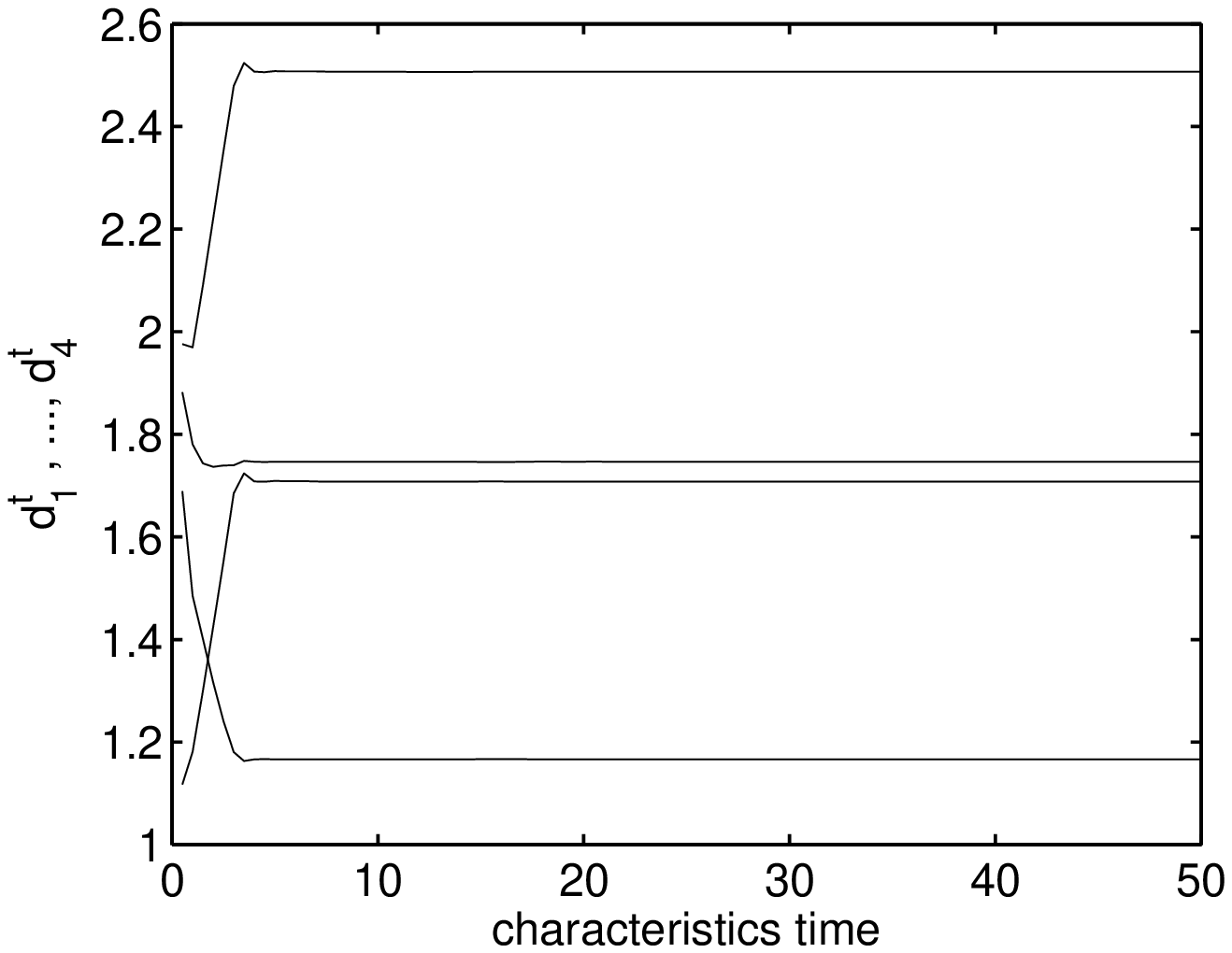}}
\subfloat[]{
\includegraphics[width=1.5in]{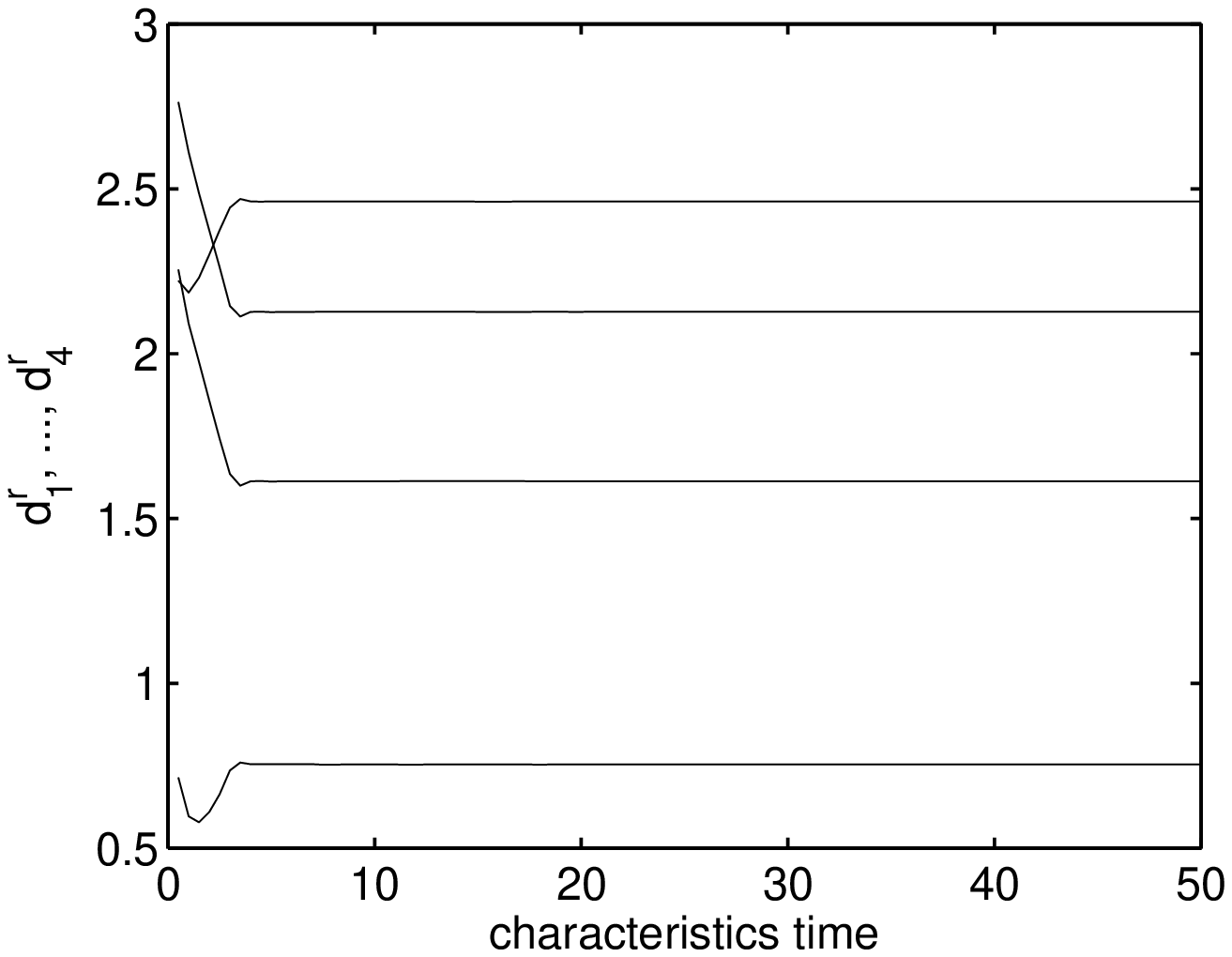}}
\hfil
\subfloat[]{
\includegraphics[width=1.5in]{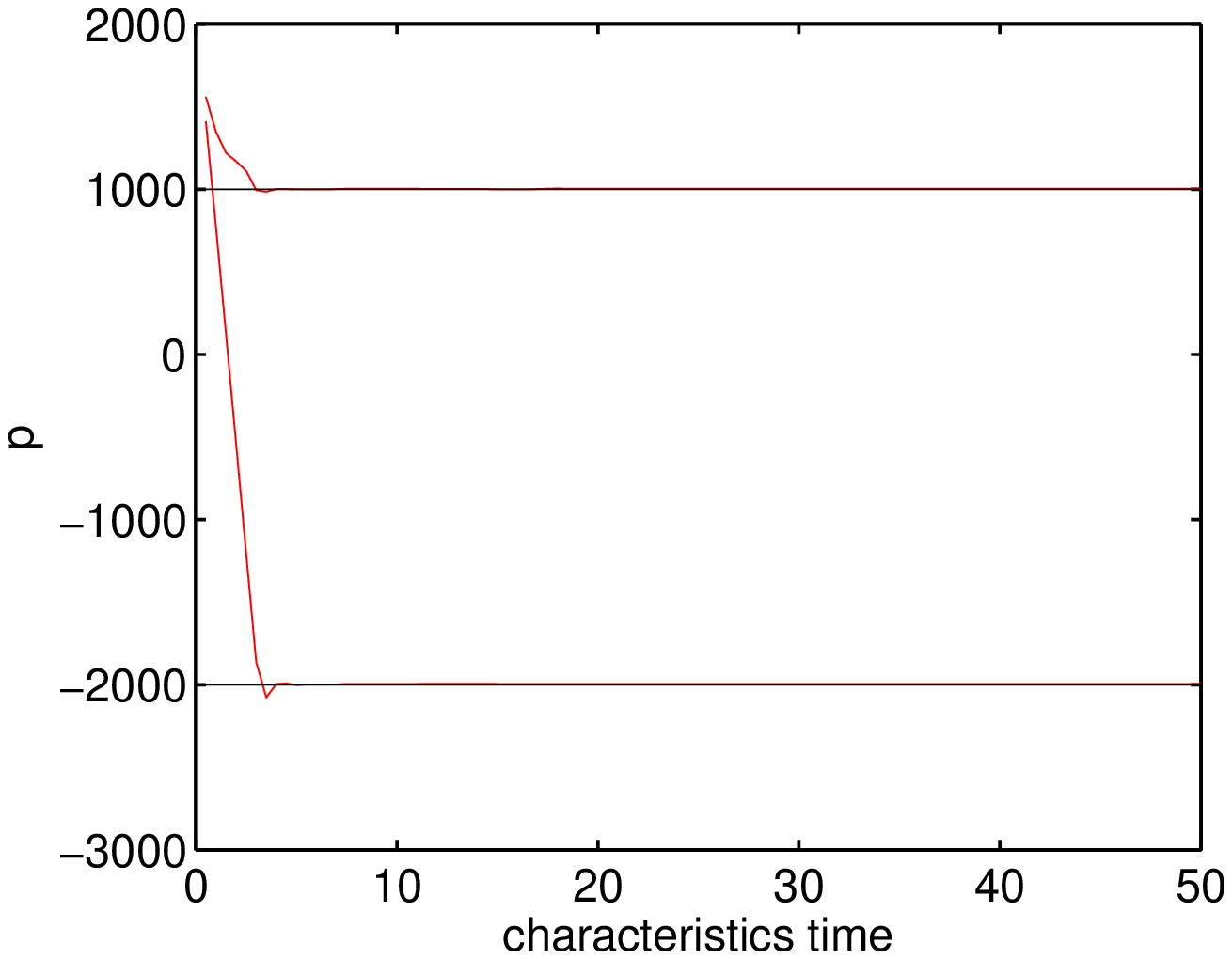}}
\subfloat[]{
\includegraphics[width=1.5in]{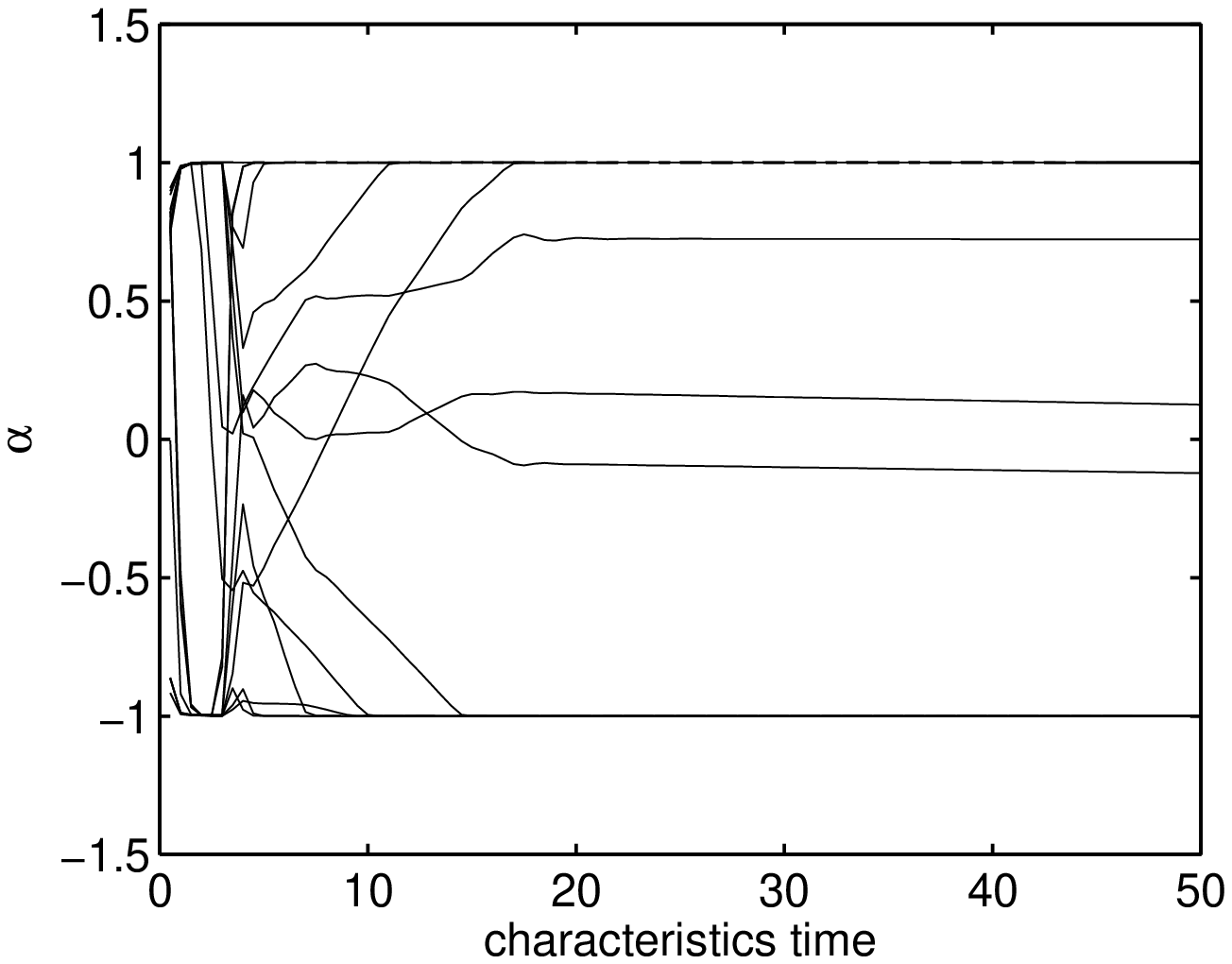}}
\hfil
\subfloat[]{
\includegraphics[width=1.5in]{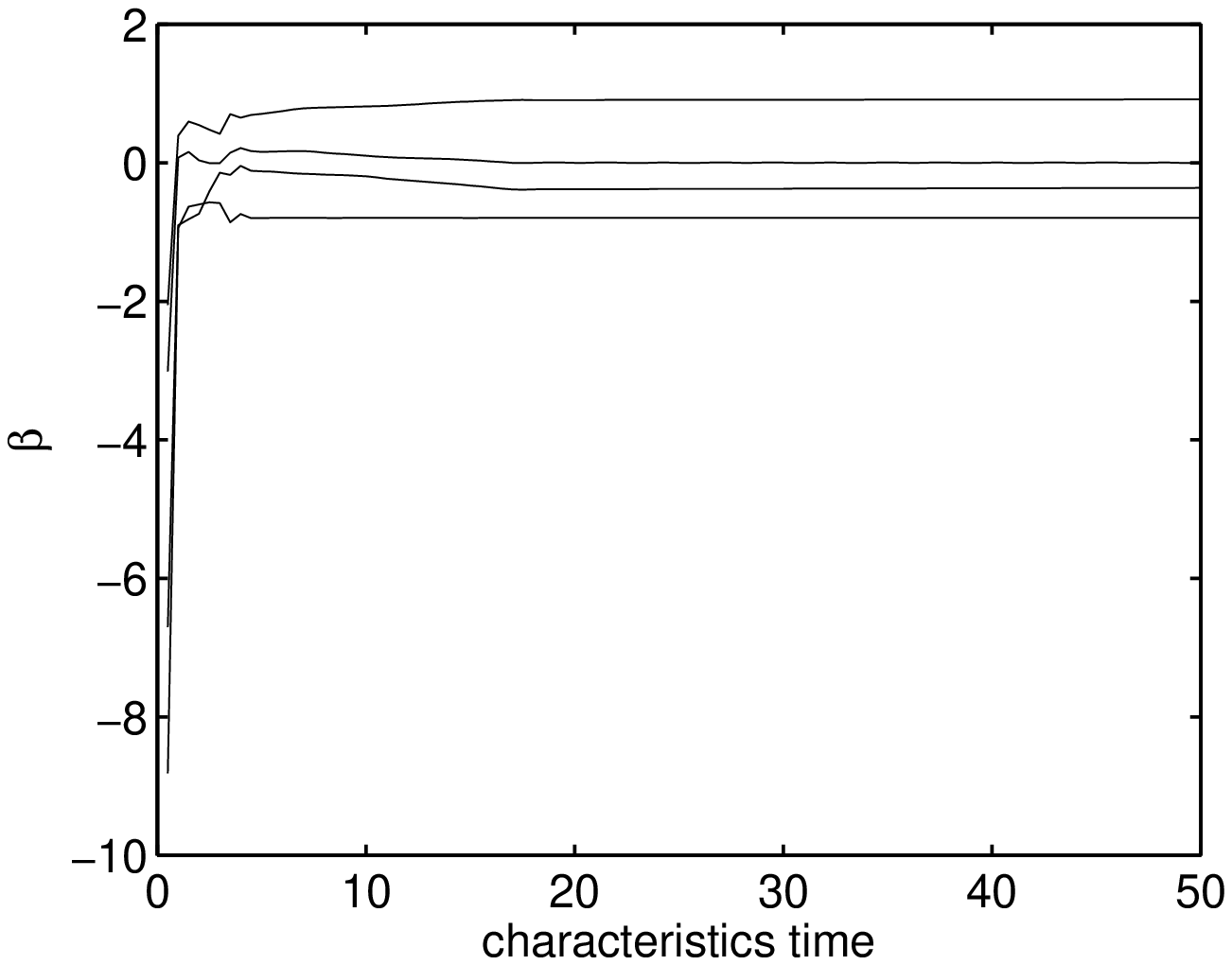}}
\subfloat[]{
\includegraphics[width=1.5in]{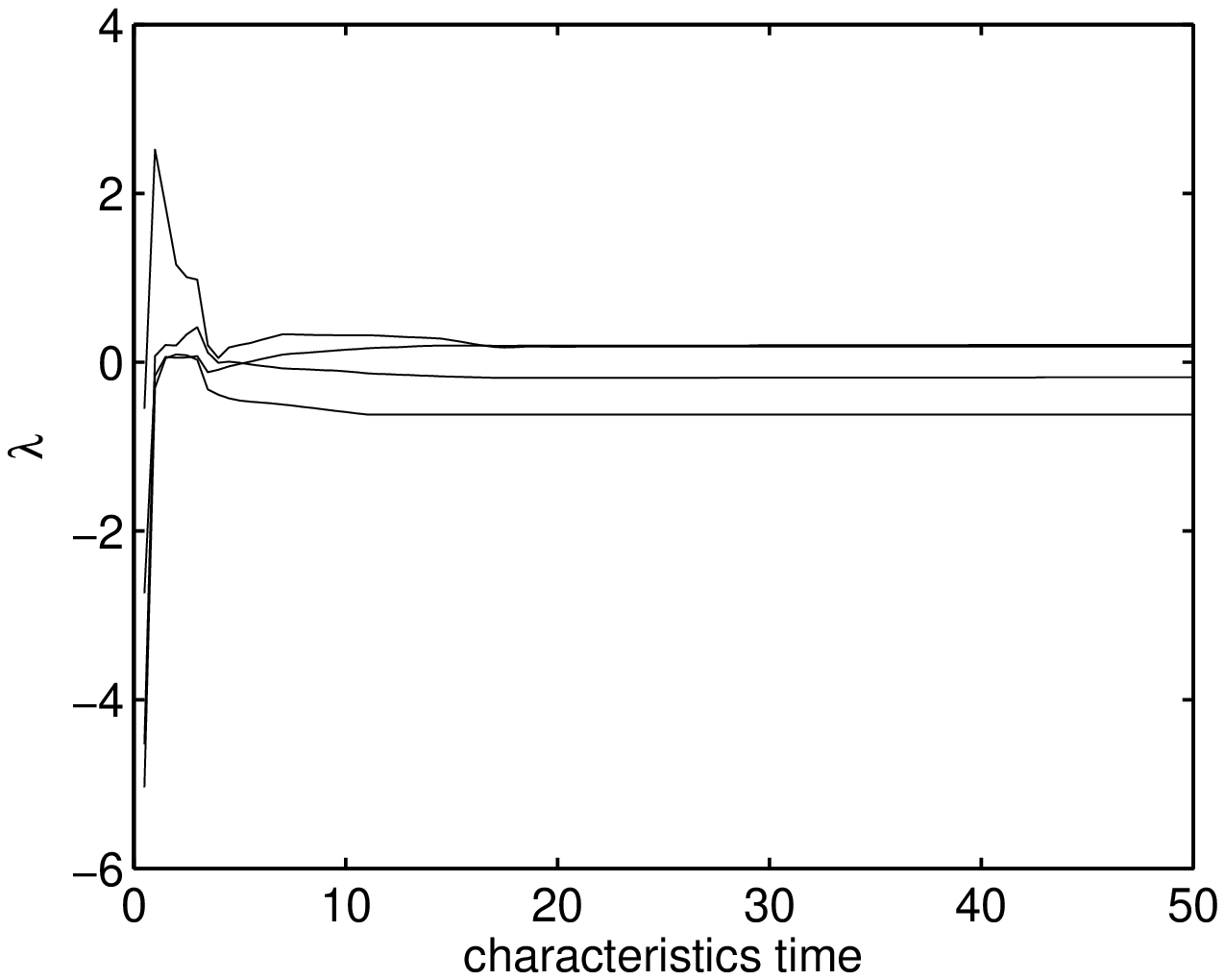}}
\caption{Dynamics of estimated parameters when the variance of Gaussian noise is $100$ (m$^2$), the standard deviation of outlier is $1000$ (m).
(a) $\ibu$; (b) $\ibz$; (c) $d_1^t, ..., d_4^t$; (d) $d_1^r, ..., d_4^r$; (e) $\ibp$ (f) $\bal$; (g) $\beta_1, ..., \beta_4$ ; (h) $\lambda_1, ..., \lambda_4$.}
\label{fig-dynamics}
\end{figure}

A typical example of the dynamics of the second method is given in Fig.~\ref{fig-dynamics}. The settings are described in the second experiment of Section~\ref{section5}. From Fig.~\ref{fig-dynamics}, we see that the network can settle down within around $50$ characteristic times.

\section{Local Stability of Proposed Algorithms}\label{section4}
In this section, we prove the local stability of the proposed methods. Local stability means that a minimum point should be stable.
Otherwise, the network never converges to the minimum.

For method 1, we let $\ibx=\left[{\ibz}^\mathrm{T}, {\ibp}^\mathrm{T}, {d^t_1}, \cdots, {d^t_M}, {d^r_1}, \cdots, {d^r_N}\right]^\mathrm{T}$ be the decision variable vector and $\{\ibx^*, \bal^*, \boldsymbol{\beta}^*, \blam^*\}$ be a minimum point of the dynamics given by \eqref{method1dynamic1_1}-\eqref{method1dynamic1_7}.
According to Theorem 1 in~\cite{lpnn0}, there are two sufficient conditions for local stability in the LPNN approach.
The first one is convexity, i.e., the Hessian matrix of the Lagrangian at $\{\ibx^*, \bal^*, \boldsymbol{\beta}^*, \blam^*\}$ should be positive definite.
It has been achieved by introducing the augmented terms.
When $C$ is large enough, at the minimum point the Hessian matrix is positive definite under mild conditions~\cite{lpnn0,lpnn1,lpnn2,lpnn3,lpnn4}.

The second one is that at the minimum point, the gradient vectors of the constraints with respect to the decision variable vector should be linearly independent.
In our case, we have $MN+M+N$ constraints, namely,
\begin{eqnarray}
h_{M(j-1)+i}(\ibp,\ibz,d^t_i,d^r_j)= z_{M(j-1)+i}-\hat{d}_{i,j}-d^t_i-d^r_j  \label{eq:constraints1}\\
h_{MN+i}(\ibp,\ibz,d^t_i,d^r_j)= {d^t_i}^2-\|\ibp-\ibt_i\|_2^2,\label{eq:constraints2}\\
h_{MN+M+j}(\ibp,\ibz,d^t_i,d^r_j)= {d^r_j}^2-\|\ibp-\ibr_j\|_2^2, \label{eq:constraints3}
\end{eqnarray}
where $i=1,\cdots,M$ and $j=1,\cdots,N$. The gradient vectors of these constraints at the minimum point are given by
\begin{eqnarray}
\left\{\!
\left[\!
  \begin{array}{c}
    \displaystyle \frac{\partial h_1(\ibx^*)}{\partial \ibp} \\
    \displaystyle \frac{\partial h_1(\ibx^*)}{\partial \ibz} \\
    \displaystyle \frac{\partial h_1(\ibx^*)}{\partial d_1^t} \\
    \displaystyle \vdots \\
    \displaystyle \frac{\partial h_1(\ibx^*)}{\partial d_M^t} \\
    \displaystyle \frac{\partial h_1(\ibx^*)}{\partial d_1^r} \\
    \displaystyle \vdots \\
    \displaystyle \frac{\partial h_1(\ibx^*)}{\partial d_N^r} \\
  \end{array}
\!\right],\!\displaystyle \cdots\!,
\left[\!
  \begin{array}{c}
    \displaystyle \frac{\partial h_{MN+M+N}(\ibx^*)}{\partial \ibp} \\
    \displaystyle \frac{\partial h_{MN+M+N}(\ibx^*)}{\partial \ibz} \\
    \displaystyle \frac{\partial h_{MN+M+N}(\ibx^*)}{\partial d_1^t} \\
    \displaystyle \vdots \\
    \displaystyle \frac{\partial h_{MN+M+N}(\ibx^*)}{\partial d_M^t} \\
    \displaystyle \frac{\partial h_{MN+M+N}(\ibx^*)}{\partial d_1^r} \\
    \displaystyle \vdots \\
    \displaystyle \frac{\partial h_{MN+M+N}(\ibx^*)}{\partial d_N^r} \\
  \end{array}
\!\right]
\!\right\} \nonumber
\end{eqnarray}
\begin{eqnarray}
={
\overbrace{\left[\!\!\!
  \begin{array}{c}
    \mathbf{0} \\
    1 \\
    0 \\
    \vdots \\
    0 \\
    1 \\
    0 \\
    \vdots \\
    0 \\
    1 \\
    0 \\
    \vdots \\
    0 \\
  \end{array}
\!\!\!\right]\!\!\!,\!\!\!\cdots\!\!\!,
\!\!\!\left[\!\!\!
  \begin{array}{c}
    \mathbf{0} \\
    0 \\
    0 \\
    \vdots \\
    1 \\
    0 \\
    0 \\
    \vdots \\
    1 \\
    0 \\
    0 \\
    \vdots \\
    1 \\
  \end{array}
\!\!\!\right]\!\!\!}^{MN},
\overbrace{\left[\!\!\!
  \begin{array}{c}
    \ibt_1-\ibp^* \\
    0 \\
    0 \\
    \vdots \\
    0 \\
    2{d_1^t}^* \\
    0 \\
    \vdots \\
    0 \\
    0 \\
    0 \\
    \vdots \\
    0 \\
  \end{array}
\!\!\!\right]\!\!\!,\!\!\!\cdots\!\!\!,
\!\!\!\left[\!\!\!
  \begin{array}{c}
    \ibt_M-\ibp^* \\
    0 \\
    0 \\
    \vdots \\
    0 \\
    0 \\
    0 \\
    \vdots \\
    2{d_M^t}^* \\
    0 \\
    0 \\
    \vdots \\
    0 \\
  \end{array}
\!\!\!\right]\!\!\!}^{M},
\overbrace{\left[\!\!\!
  \begin{array}{c}
    \ibr_1-\ibp^* \\
    0 \\
    0 \\
    \vdots \\
    0 \\
    0 \\
    0 \\
    \vdots \\
    0 \\
    2{d_1^r}^* \\
    0 \\
    \vdots \\
    0 \\
  \end{array}
\!\!\!\right]\!\!\!,\!\!\!\cdots\!\!\!,
\!\!\!\left[\!\!\!
  \begin{array}{c}
    \ibr_N-\ibp^*\\
    0 \\
    0 \\
    \vdots \\
    0 \\
    0 \\
    0 \\
    \vdots \\
    0 \\
    0 \\
    0 \\
    \vdots \\
    2{d_N^r}^* \\
  \end{array}
\!\!\!\right]\!\!\!}^{N},
}
\end{eqnarray}
where $\mathbf{0}=[0,0]^\mathrm{T}$. We can see there are $MN+M+N$ gradient vectors, each one has $MN+M+N+2$ elements.
The first $MN$ columns are the gradient vectors of equality \eqref{eq:constraints1}.
Obviously, these columns are independent with each others.
Similarly, we can see that the $M$ middle columns are the gradient vectors of equality \eqref{eq:constraints2}, and the last $N$ columns are the gradient vectors of equality \eqref{eq:constraints3}.
Both of them are independent within themselves. For the constraints given in \eqref{eq:constraints1}, their gradients with respect to $\ibp^*$ are all equal to $0$.
While for the constraints given in \eqref{eq:constraints2} and \eqref{eq:constraints3}, as long as the position of target to be detected does not coincide with any transmitter or receiver, their gradients with respect to $\ibp^*$ cannot be zero.
Hence the first $MN$ columns are independent with the last $M+N$ columns. Besides, it is easy to notice that the middle $M$ columns are independent with the last $N$ columns.
Thus, the gradients of the constraints are linear independent, and the dynamics around a minimum point are stable.

For method 2, $\ibx=\left[{\ibu}^\mathrm{T}, {\ibp}^\mathrm{T}, {d^t_1}, \cdots, {d^t_M}, {d^r_1}, \cdots, {d^r_N}\right]^\mathrm{T}$.
We need to show that, at a local minimum point $\{\ibx^*, \bal^*, \boldsymbol{\beta}^*, \blam^*\}$, the gradient vectors of constraints are linearly independent. The proof process is basically similar with the first method, except that the decision variable $\ibz$ is replaced by $\ibu$. Thus, the gradient vectors of constraints at the minimum point are given by
\begin{eqnarray}
\left\{\!
\left[\!
  \begin{array}{c}
    \displaystyle \frac{\partial h_1(\ibx^*)}{\partial \ibp} \\
    \displaystyle \frac{\partial h_1(\ibx^*)}{\partial \ibu} \\
    \displaystyle \frac{\partial h_1(\ibx^*)}{\partial d_1^t} \\
    \displaystyle \vdots \\
    \displaystyle \frac{\partial h_1(\ibx^*)}{\partial d_M^t} \\
    \displaystyle \frac{\partial h_1(\ibx^*)}{\partial d_1^r} \\
    \displaystyle \vdots \\
    \displaystyle \frac{\partial h_1(\ibx^*)}{\partial d_N^r} \\
  \end{array}
\!\right],\!\displaystyle \cdots\!,
\left[\!
  \begin{array}{c}
    \displaystyle \frac{\partial h_{MN+M+N}(\ibx^*)}{\partial \ibp} \\
    \displaystyle \frac{\partial h_{MN+M+N}(\ibx^*)}{\partial \ibu} \\
    \displaystyle \frac{\partial h_{MN+M+N}(\ibx^*)}{\partial d_1^t} \\
    \displaystyle \vdots \\
    \displaystyle \frac{\partial h_{MN+M+N}(\ibx^*)}{\partial d_M^t} \\
    \displaystyle \frac{\partial h_{MN+M+N}(\ibx^*)}{\partial d_1^r} \\
    \displaystyle \vdots \\
    \displaystyle \frac{\partial h_{MN+M+N}(\ibx^*)}{\partial d_N^r} \\
  \end{array}
\!\right]
\!\right\} \nonumber
\end{eqnarray}
\begin{eqnarray}
={
\overbrace{\left[\!\!\!
  \begin{array}{c}
    \mathbf{0} \\
    g_{1} \\
    0 \\
    \vdots \\
    0 \\
    1 \\
    0 \\
    \vdots \\
    0 \\
    1 \\
    0 \\
    \vdots \\
    0 \\
  \end{array}
\!\!\!\right]\!\!\!,\!\!\!\cdots\!\!\!,
\!\!\!\left[\!\!\!
  \begin{array}{c}
    \mathbf{0} \\
    0 \\
    0 \\
    \vdots \\
    g_{MN} \\
    0 \\
    0 \\
    \vdots \\
    1 \\
    0 \\
    0 \\
    \vdots \\
    1 \\
  \end{array}
\!\!\!\right]\!\!\!}^{MN},
\overbrace{\left[\!\!\!
  \begin{array}{c}
    \ibt_1-\ibp^* \\
    0 \\
    0 \\
    \vdots \\
    0 \\
    2{d_1^t}^* \\
    0 \\
    \vdots \\
    0 \\
    0 \\
    0 \\
    \vdots \\
    0 \\
  \end{array}
\!\!\!\right]\!\!\!,\!\!\!\cdots\!\!\!,
\!\!\!\left[\!\!\!
  \begin{array}{c}
    \ibt_M-\ibp^* \\
    0 \\
    0 \\
    \vdots \\
    0 \\
    0 \\
    0 \\
    \vdots \\
    2{d_M^t}^* \\
    0 \\
    0 \\
    \vdots \\
    0 \\
  \end{array}
\!\!\!\right]\!\!\!}^{M},
\overbrace{\left[\!\!\!
  \begin{array}{c}
    \ibr_1-\ibp^* \\
    0 \\
    0 \\
    \vdots \\
    0 \\
    0 \\
    0 \\
    \vdots \\
    0 \\
    2{d_1^r}^* \\
    0 \\
    \vdots \\
    0 \\
  \end{array}
\!\!\!\right]\!\!\!,\!\!\!\cdots\!\!\!,
\!\!\!\left[\!\!\!
  \begin{array}{c}
    \ibr_N-\ibp^*\\
    0 \\
    0 \\
    \vdots \\
    0 \\
    0 \\
    0 \\
    \vdots \\
    0 \\
    0 \\
    0 \\
    \vdots \\
    2{d_N^r}^* \\
  \end{array}
\!\!\!\right]\!\!\!}^{N},
}
\end{eqnarray}
where $k=M(j-1)+i$,
\beq
g_{k} &=& \frac{\partial h_{k}(\ibx^*)}{\partial z_{k}}\frac{\partial z_{k}}{\partial u_{k}}
=\frac{1}{1 + \exp{(-\eta (|u_{k}| - 1))}} \nonumber \\
&&+ \frac{\eta(|u_{k}|-\delta)\exp{(-\eta (|u_{k}| - 1))}}{(1 + \exp{(-\eta (|u_{k}| - 1))})^2}. \nonumber
\eeq
For the case with $L_1$-norm objective function, $\eta\rightarrow \infty$, $\delta=1$.
If we assume $z_i\neq 0$, in other words, all data points are influenced by noise, thus we have $g_k=1$ for $\forall k=1,\dots,MN$. When the proximate $L_0$-norm objective function is used, we let $\eta$ be a large positive number, $\delta=0$. Without any assumption, we can deduce that, for $\forall k=1,\dots,MN$, $g_k$ is a positive constant. Next, similar with the proof process of method 1, we can prove that a minimum point of the second method has local stability.

\section{Numerical Examples}\label{section5}
\begin{figure}[ht]
\centering
\centerline{\includegraphics[width=3.2in]{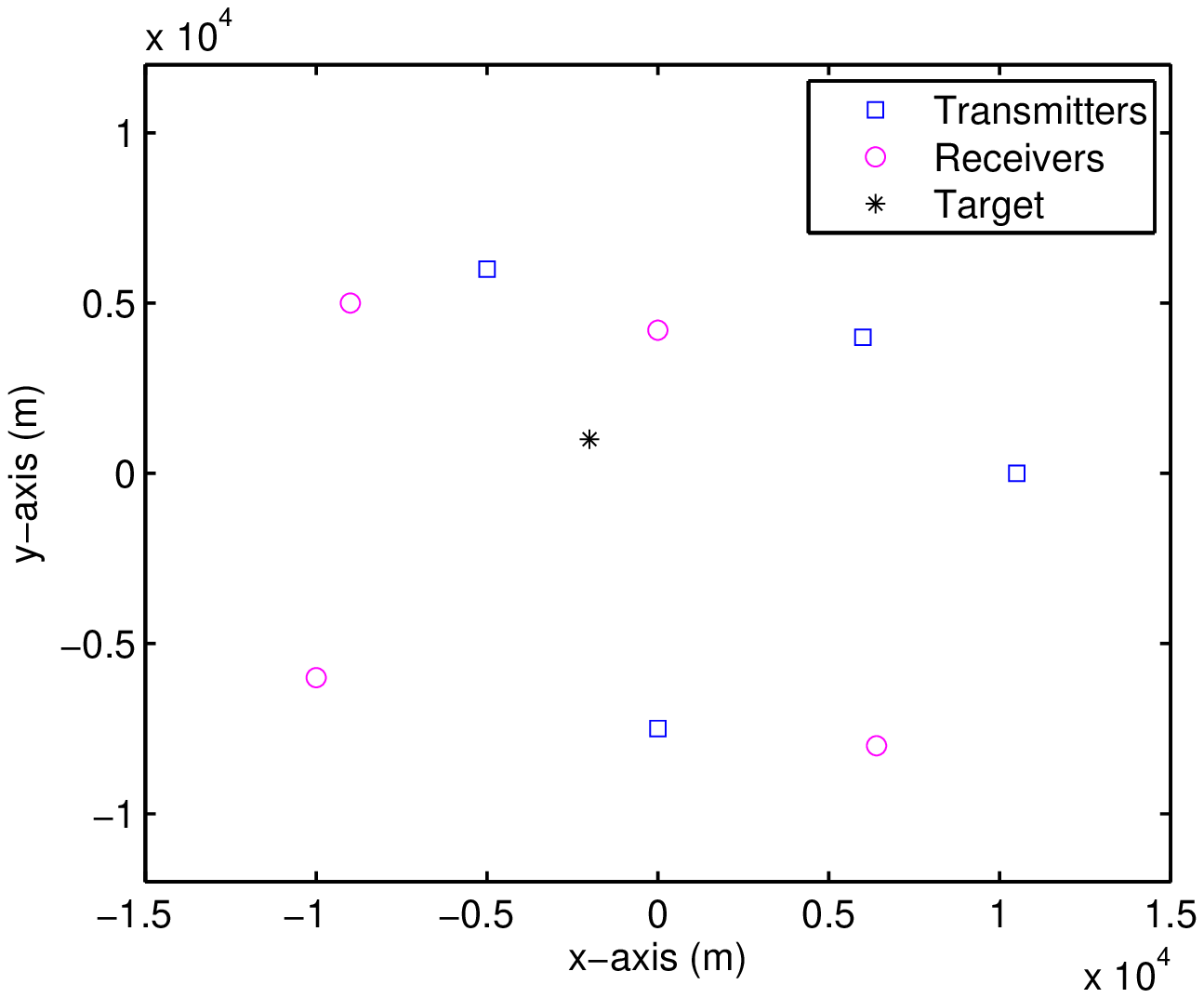}}
\caption{The configuration of transmitters, receivers and target.}
\label{fig-system}
\end{figure}

In this section, we conduct several simulations and experiments to test the performance of our proposed algorithms. First, we discuss the parameter settings and the initialization. In our MIMO radar localization system, $4$ transmitters and $4$ receivers are used, i.e., $M=N=4$. Their positions are $\ibt_1=[-5000,6000]^\mathrm{T} {\rm m }, \ibt_2=[0,7500]^\mathrm{T} {\rm m }, \ibt_3=[10500,0]^\mathrm{T} {\rm m }, \ibt_4=[6000,4000]^\mathrm{T} {\rm m }$ and $\ibr_1=[-10000,-6000]^\mathrm{T} {\rm m }, \ibr_2=[-9000,5000]^\mathrm{T} {\rm m }, \ibr_3=[0,4200]^\mathrm{T} {\rm m }, \ibr_4=[6400,-8000]^\mathrm{T} {\rm m }$, respectively. The true position of the target is $\ibp=[-2000,1000]^\mathrm{T} {\rm m }$. The geometry of transmitters, receivers and target are shown in Fig.~\ref{fig-system}.

For the proposed approaches, we set $C=20$ and step sizes $\mu_1=\mu_4=\mu_5=\mu_6=\mu_7=10^{-3}, \mu_2=\mu_3=10^{-5}$. The initial values of variables $\ibp, \ibz, d^t_i(i=1, \cdots ,M), d^r_j(j=1, \cdots ,N), \bal=[\al_1,...,\al_{MN}], \boldsymbol{\beta}=[\beta_1,...,\beta_M], \boldsymbol{\lam}=[\lam_1,...,\lam_N]$ are some small random values. And we set $a=50$ in the first method.
Two state-of-the-art algorithms are implemented for performance comparison. They are the target localization algorithm described in \cite{lpnn4} and the robust target localization algorihtm given by \cite{liang2016robust}.
The former is also based on LPNN framework, but it assumes that the noise follows Gaussian distribution and uses $l_2$-norm in its objective function. While, the latter is a robust target localization algorithm for MIMO radar system.
It introduces the maximum correntropy criterion (MCC) into the conventional ML method to deal with outliers, and apply half-quadratic optimization technique to handle the corresponding nonconvex nonlinear function.
\subsection{Experiment 1: Target Localization in Gaussian Noise}
\begin{figure}[!h]
\centering
\centerline{\includegraphics[width=3.2in]{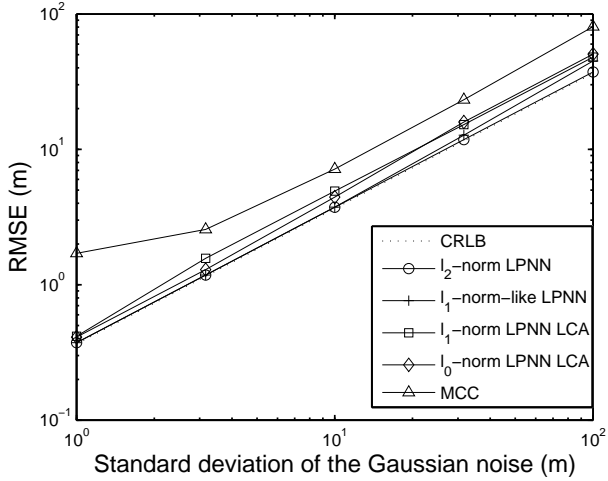}}
\caption{The RMSE results of different algorithms. The standard deviation of the Gaussian noise ranges from $1$ to $10^2$.}
\label{fig-ex1}
\end{figure}

In the first experiment we test the root mean squared error (RMSE) performance of the proposed algorithms under Gaussian noise environment without introducing any outliers. The standard deviation of the Gaussian noise varies from $1$ to $10^2$. For each noise level, we repeat the experiment 100 times. The results are shown in Fig.~\ref{fig-ex1}.

In Fig.~\ref{fig-ex1}, CRLB is short for Cram$\acute{e}$r-Rao low bound, which denotes a lower bound on the variance of any unbiased estimator
\cite{lpnn4}; $l_2$-norm LPNN denotes the algorithm given by \cite{lpnn4}; $l_1$-norm-like LPNN is our proposed method $1$; $l_1$-norm LPNN LCA and $l_0$-norm LPNN LCA represent our proposed method $2$ with $l_1$-norm objective function and $l_0$-norm objective function respectively; MCC is the robust algorithm given in \cite{liang2016robust}. In Fig.~\ref{fig-ex1},  we see that the performance of our proposed two robust algorithms is closed to the CRLB in Gaussian noise environment. They are better than the robust algorithm given by \cite{liang2016robust} and are slightly inferior to the algorithm given by \cite{lpnn4}, which is based on the Gaussian noise model.

\subsection{Experiment 2: Target Localization in Gaussian Noise with NLOS Outliers}
\begin{figure}[!h]
\centering
\centerline{\includegraphics[width=3.2in]{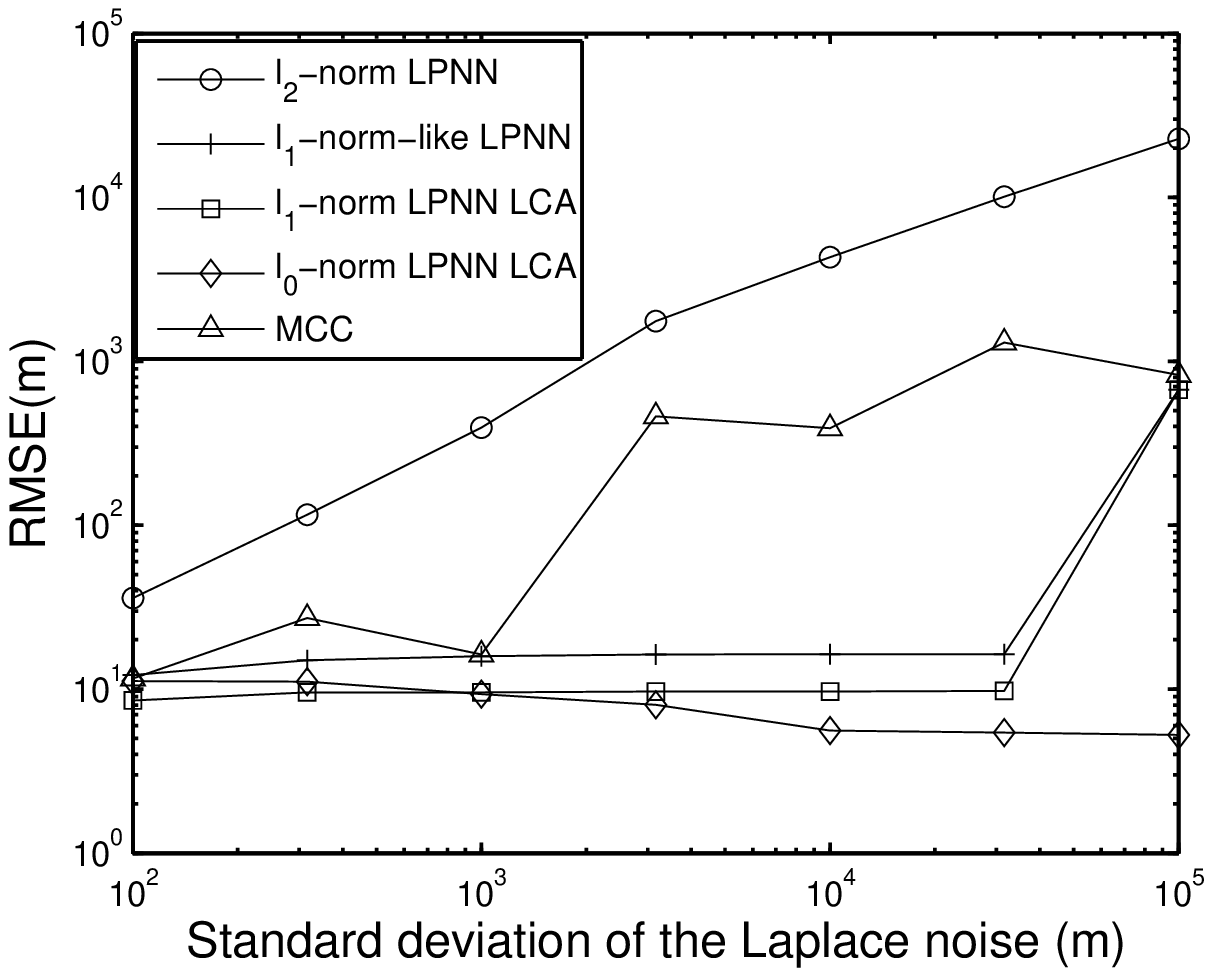}}
\caption{The RMSE results of different algorithms. The standard deviation of the exponential distribution (outlier level) is varied from $10^2$ m to $10^{5}$ m.}
\label{fig-ex2_1}
\end{figure}

The basic setting of the second experiment is similar with the first one, but we fix the variance of Gaussian noise to $100$, and introduce outliers into the measurement matrix. We assume that there exists non-line-of-sight (NLOS) propagation between one transmitter and the target or between the target and one receiver. Thus all measurements associated with this transmitter or receiver include NLOS outliers \cite{nouvel2014study,setlur2014multipath}.
In this experiment, we randomly choose one of the transmitters or receives and add NLOS outliers into its relevant measured values.
In other words, the measurements can be seen as a $4\times 4$ matrix, one of columns or rows is influenced by NLOS outliers.
The outliers are generated by exponential distribution.
Then we conduct the experiments under different outlier levels.
For each different outlier level, we also repeat the experiment $100$ times.  The results are shown in Fig.~\ref{fig-ex2_1}.

From Fig.~\ref{fig-ex2_1}, we see that our first method and second method with $l_1$-norm objective have decent performance when the outlier level (standard deviation of the exponential distribution) is less than $10^{4.5}$. Both of them break down when the outlier level is $10^{5}$. While the second method with $l_0$-norm objective function may not handle the low level outliers very well, but it is very effective to reduce the influence high level
outliers. Conversely, the algorithm given by \cite{lpnn4} is very sensitive with outliers. For the robust target localization algorithm \cite{liang2016robust}, even though it can also effectively reduce the influence of outliers, its performance is worse than our proposed methods.

\begin{figure}[!h]
\centering
\centerline{\includegraphics[width=3.2in]{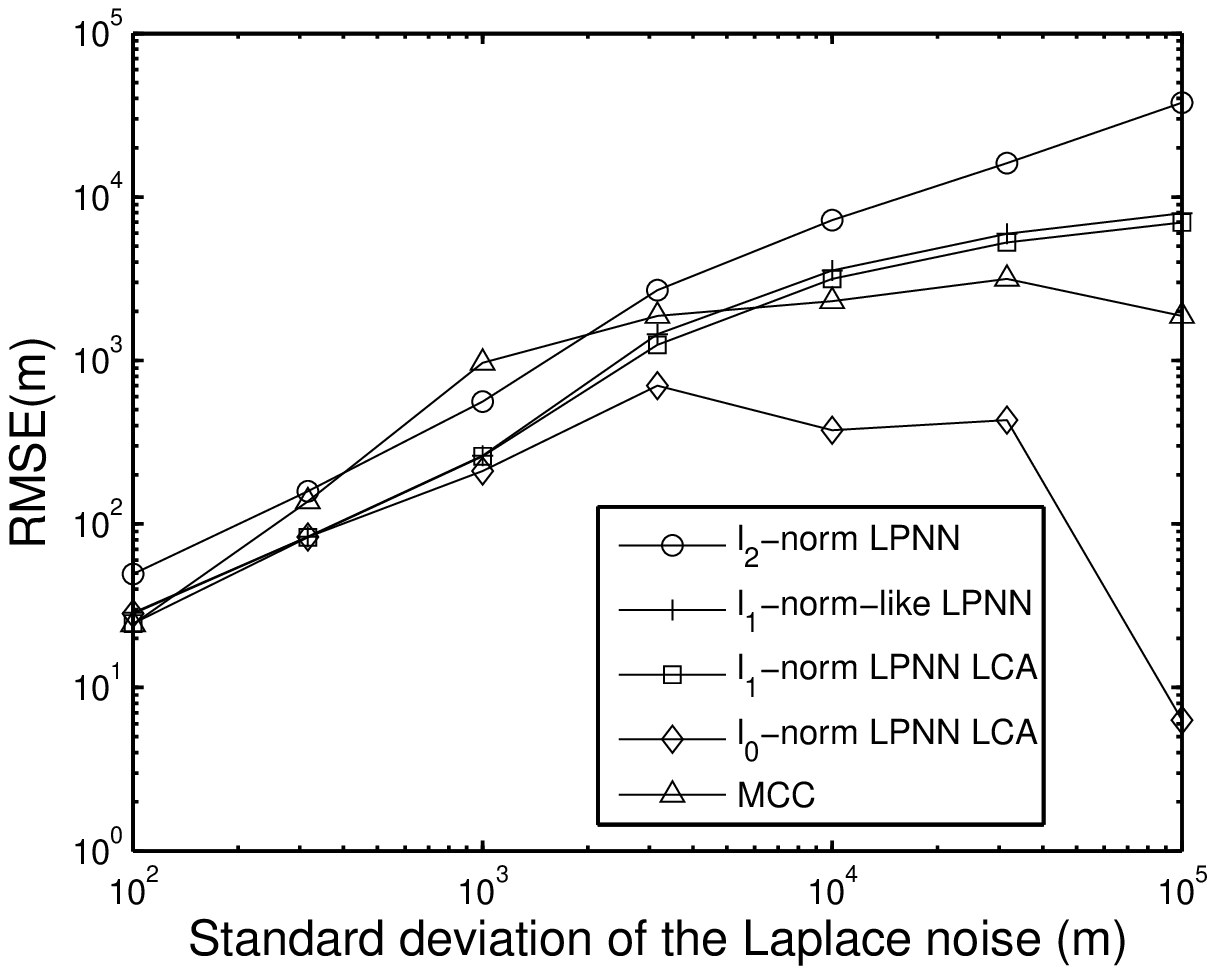}}
\caption{The RMSE results of different algorithms. The standard deviation of the exponential distribution (outlier level) is varied from $10^2$ m to $10^{5}$ m.}
\label{fig-ex2_2}
\end{figure}

In the following experiment, we randomly choose one transmitter and one receiver, and then add NLOS outliers into their relevant measurements. Thus $7$ elements in the $4\times 4$ measurement matrix include outliers. The results are shown in Fig.~\ref{fig-ex2_2}.

\begin{figure}[!h]
\centering
\centerline{\includegraphics[width=3.2in]{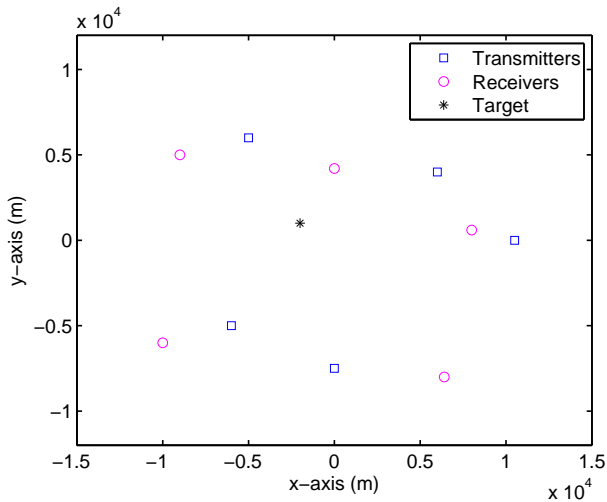}}
\caption{The configuration of transmitters, receivers and target.}
\label{fig-system_2}
\end{figure}

From Fig.~\ref{fig-ex2_2}, we see that the robust target localization algorithm given by \cite{liang2016robust} and our proposed algorithms can reduce the influence of outliers in this case. However, due to the high proportion of outliers, all mentioned algorithms cannot give satisfied results. But among them, the performance of $l_0$-norm LPNN LCA is still the best. Then, we further add one transmitter $\ibt_5=[-6000,-5000]^\mathrm{T} m$ and one receiver $\ibr_5=[8000,600]^\mathrm{T} m$. After that, the geometry of transmitters and receivers is depicted in Fig.~\ref{fig-system_2}.

\begin{figure}[!h]
\centering
\centerline{\includegraphics[width=3.2in]{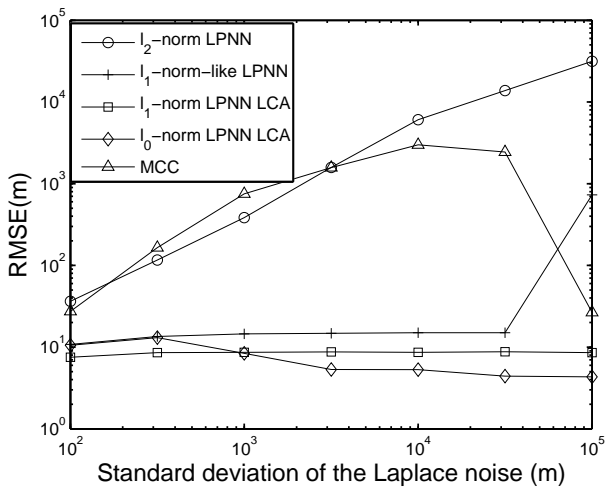}}
\caption{The RMSE results of different algorithms. The standard deviation of the exponential distribution (outlier level) is varied from $10^2$ m to $10^{5}$ m.}
\label{fig-ex2_3}
\end{figure}

Then, we still randomly choose one transmitter and one receiver, add NLOS outliers into their relevant measurements. Thus $9$ elements in the $5\times 5$ measurement matrix are with outliers. The results are shown in Fig.~\ref{fig-ex2_3}.

From Fig.~\ref{fig-ex2_3}, we see the performance of our proposed algorithms is better than others. And, generally speaking, $l_0$-norm LPNN LCA is the best.
\subsection{Experiment 3: Target Localization in Gaussian Noise with Some SINR Outliers}
\begin{figure}[!h]
\centering
\centerline{\includegraphics[width=3.2in]{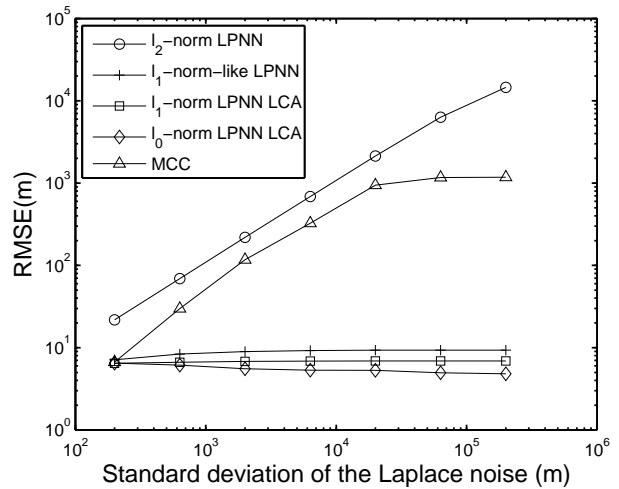}}
\caption{The RMSE results of different algorithms under SINR outliers. The variance of Gaussian noise is equal to 100. The standard deviation of the Laplace noise ranges from $2\times 10^2$ m to $2\times 10^5$ m.}
\label{fig-ex3}
\end{figure}

In the third experiment, we evaluate the performance of our proposed algorithms under low signal-to-interference-noise ratio (SINR) environment. This experiment is implemented based on the MIMO radar system given by Fig.~\ref{fig-system}. First, we set the variance of Gaussian noise to $100$ and introduce outliers, which are generated to model the low SINR environment. 
Assume that 5 measurement values $\hat{d}_{1,2}$, $\hat{d}_{2,3}$, $\hat{d}_{3,4}$, $\hat{d}_{4,1}$, and $\hat{d}_{1,4}$, are influenced by SINR outliers.
The SINR outliers can be both negative and positive, they are generated by Laplace distribution in this experiment. The standard deviation of Laplace distribution ranges from $2\times10^2$ (m) to $2\times10^5$ (m). At each outlier level, we still repeat the experiment 100 times. The results are shown in Fig.~\ref{fig-ex3}.

From Fig.~\ref{fig-ex3}, we see that under low SINR environment, the performance of our proposed algorithms is also superior to others. The performance of $l_0$-norm LPNN LCA is still the best.



\section{Conclusion}\label{section6}
In this paper, two algorithms for solving target localization problem in a distributed MIMO radar system were developed. To alleviate the influence of outliers, we utilize the property of $l_p$-norm ($p=1$, or $p=0$) and redesign the objective function of the original optimization problem.
To achieve a real-time solution, both two proposed algorithms are based on the concept of LPNN. Since the objective function of the proposed mathematic model is non-differentiable. Two approaches are devised for solving this problem. In first method, we use a differentiable function to approximate the $l_1$-norm in objective function. While, in the second method, we combine the LCA with LPNN framework, and propose an algorithm which not only solves the non-differentiable problem of $l_1$-norm but also $l_0$-norm. The experiments demonstrate that the proposed algorithms can effectively reduce the influence of outliers and they are superior to several state-of-the-art MIMO radar target localization methods.


\appendices

\ifCLASSOPTIONcaptionsoff
  \newpage
\fi



%
%
\bibliographystyle{IEEEtran}
\bibliography{my_reference}

\end{document}